\documentclass[aip,amsmath,amssymb,floatfix,preprint,citeautoscript,noeprint,superscriptaddress]{revtex4-2}

\usepackage{lipsum} 
\usepackage{bibentry}
\usepackage[english]{babel}
\usepackage[usenames,dvipsnames]{xcolor}
\usepackage{graphicx}
\usepackage[caption=false]{subfig} 
\usepackage{amsmath,amssymb,bm}
\usepackage[version=3]{mhchem}
\usepackage{verbatim}
\usepackage{multirow}
\usepackage{dcolumn}
\usepackage{float}
\usepackage{nicefrac}
\usepackage{siunitx}
\usepackage{booktabs}
\usepackage{chemformula}
\usepackage{wrapfig}
\usepackage{enumitem}  
\usepackage{transparent}
\usepackage[colorlinks,allcolors=black,citecolor=blue,urlcolor=blue]{hyperref}
\usepackage{tcolorbox}
\usepackage{ulem}

\DeclareSIUnit[number-unit-product = {\,}]{\amu}{amu}
\DeclareSIUnit[number-unit-product = {\,}]{\kJmol}{\kilo\joule\per\mol}
\DeclareSIUnit[number-unit-product = {\,}]{\Nsm}{\newton\second\per\meter\cubed}
\DeclareSIUnit[number-unit-product = {\,}]{\THz}{\tera\hertz}
\DeclareSIUnit[number-unit-product = {\,}]{\meV}{\milli\electronvolt}
\DeclareSIUnit[number-unit-product = {\,}]{\cal}{cal}

\newcommand{\tcr}[1]{\textcolor{black}{#1}}
\newcommand{\tcb}[1]{\textcolor{black}{#1}}

\newcommand{\rr}{\mathbf{r}}
\renewcommand{\d}{\mathrm{d}}

\newcommand{\rev}[1]{{\color{black} {#1}}}

\newcommand\tb[1]{\textbf{#1}}

\newcommand\mc[1]{\mathcal{#1}}
\newcommand\e[1]{\cdot 10^{#1}}
\newcommand\dd{\textnormal{d}}
\newcommand\beq{\begin{equation}}
\newcommand\eeq{\end{equation}}
\newcommand\beqa{\begin{eqnarray}}
\newcommand\eeqa{\end{eqnarray}}

\newcommand\im[1]{\textnormal{Im}\left[#1\right]}

\def\q{\tb{q}}
\def\v{\tb{v}}

\def\A{\mc{A}}

\usepackage{sansmath}
\sansmath

\begin{document}

\title{\large{\rev{Momentum tunneling between nanoscale liquid flows}}}

\author{Baptiste Coquinot$^{1,2\dagger}$, Anna T. Bui$^{3,\dagger}$, Damien Toquer$^{1}$,  Angelos Michaelides$^3$, Nikita Kavokine$^{2,4,5*}$, Stephen J. Cox$^{3,6*}$ and Lyd\'eric Bocquet$^{1*}$}
\email{sjc236@cam.ac.uk, nikita.kavokine@epfl.ch, lyderic.bocquet@ens.fr}
\affiliation{\scriptsize Laboratoire de Physique de l'Ecole Normale Sup\'erieure, 24 rue Lhomond, 75005, Paris, France}
\affiliation{\scriptsize Max Planck Institute for Polymer Research, Ackermannweg 10, Mainz, Germany}
\affiliation{\scriptsize Yusuf Hamied Department of Chemistry, University of Cambridge, Cambridge CB2 1EW, United Kingdom}
\affiliation{\scriptsize Center for Computational Quantum Physics, Flatiron Institute, 162 $5^{\rm th}$ Avenue, 10010 New York, USA }
\affiliation{\scriptsize The Quantum Plumbing Lab (LNQ), \'Ecole Polytechnique F\'ed\'erale de Lausanne (EPFL), Station 6, CH-1015 Lausanne, Switzerland}
\affiliation{\scriptsize Department of Chemistry, Durham University, Durham, UK \\
$\dagger$: these authors contributed equally}

\begin{abstract}\bf \small
The world of nanoscales in fluidics is the frontier where the continuum of fluid mechanics meets the atomic, and even quantum, nature of matter. 
While water dynamics remain largely classical under extreme confinement, several experiments have recently reported coupling between water transport and the electronic degrees of freedom of the confining materials. This avenue prompts us to reconsider nanoscale hydrodynamic flows under the perspective of interacting excitations, akin to condensed matter frameworks.
Here, using a combination of many-body theory and molecular simulations, we show that the flow of a liquid can induce the flow of another liquid behind a separating wall, at odds with the prediction of continuum hydrodynamics. 
We further show that the range of this ``flow tunneling'' can be tuned through the solid's electronic excitations, with a maximum occurring when these are at resonance with the 
\tcb{liquid's charge density fluctuations}. 
Flow tunneling is expected to play a role in global transport across nanoscale fluidic networks, such as lamellar graphene oxide or MXene membranes. It further suggests exploiting the electronic properties of the confining walls for manipulating liquids via their dielectric spectrum, beyond the nature and characteristics of individual molecules.
\end{abstract}

\maketitle


Nature does many exquisite things with water and ions at small scales. This stunning observation is a source of inspiration, and a strong motivation to explore fluidic transport in nanometric confinement. And, indeed, over the last ten years, a cabinet of curiosities of unconventional nanoscale flow properties has been unveiled in nanofluidic studies \cite{Aluru2023fluids,Faucher2019, Bocquet2020}.  This prompted many to revisit the 
standard frameworks of fluid dynamics. \rev{While confining walls are merely considered as boundary conditions for hydrodynamics, they are actually `jiggling and wiggling' matter, being themselves the locus of fluctuations and excitations such as phonons\cite{Ma2015,Marbach2018,Lizee2024}, plasmons\cite{Kavokine2022,Bui2023}, etc.}
In particular, while the dynamics of liquid water are essentially classical at the molecular scale---grounding our  understanding of water transport in classical physics---the confining surfaces may host delocalized electrons, whose behavior should be described within quantum mechanics. Many experimental studies have now hinted at a non-trivial coupling between the classical water dynamics and the quantum dynamics of these electrons. 
Prominent examples include: flow-induced electronic currents \cite{Lizee2023,Coquinot2023,Rabinowitz2020,Yin2014}, the modification of liquid wetting by substrate metallicity \cite{Comtet2017}, heat transfer from graphene electrons to the fluid environment \cite{Yu2023}, anomalies in hydrodynamic friction at water--carbon interfaces \cite{Majumder2005,Holt2006,Maali2008,Ma2015water,Secchi2016,Xie2018,Keerthi2021,Kavokine2022,Bui2023,Coquinot2023B} and its subtle difference with insulating materials \cite{Tocci2014}.

These findings have shifted perspectives on nanoscale hydrodynamics, prompting a departure from the traditional notion that the solid only acts as a static potential for the liquid molecules, to consider instead the liquid--solid interaction at the level of collective charge density fluctuations. Specifically, polar liquids such as water carry dielectric fluctuations from their collective intermolecular motions, spanning three orders of magnitude in the terahertz (THz) frequency range of the spectrum \cite{Heyden2010}. For carbon-based materials such as graphene and its multilayers, the THz frequency range is where low-energy electronic surface plasmon modes lie \cite{Hafez2018,Ju2011}. 
Describing water's interaction with these fluctuations is greatly simplified if its dielectric modes are formally quantized: the corresponding elementary excitations have been dubbed ``hydrons".~\cite{Coquinot2023,Yu2023}
The excitation perspective for the collective water modes---inspired by many-body condensed matter physics---is at the root of the fluctuation-induced (or ``quantum'') friction theory, which has successfully explained several of the phenomena mentioned above \cite{Kavokine2022,Bui2023,Coquinot2023,Coquinot2024} \tcb{and therefore holds the potential to reveal and explain new physics.} 

\tcb{Here, we show that, as water on one side of a solid wall is driven, the water's excited hydron modes interact with collective modes in the solid substrate. As a result, a flow is induced in the water on the other side of the wall,} at odds with the prediction of classical hydrodynamics. We dub this phenomenon ``flow tunneling". We develop a complete theoretical and numerical description of flow tunneling, elucidating the role of the solid's electronic properties in the hydron transmission process, and assessing its potential as a new principle for manipulating nanoscale liquid flows.

\begin{figure*}[t]
 \centering \includegraphics[width=\textwidth]{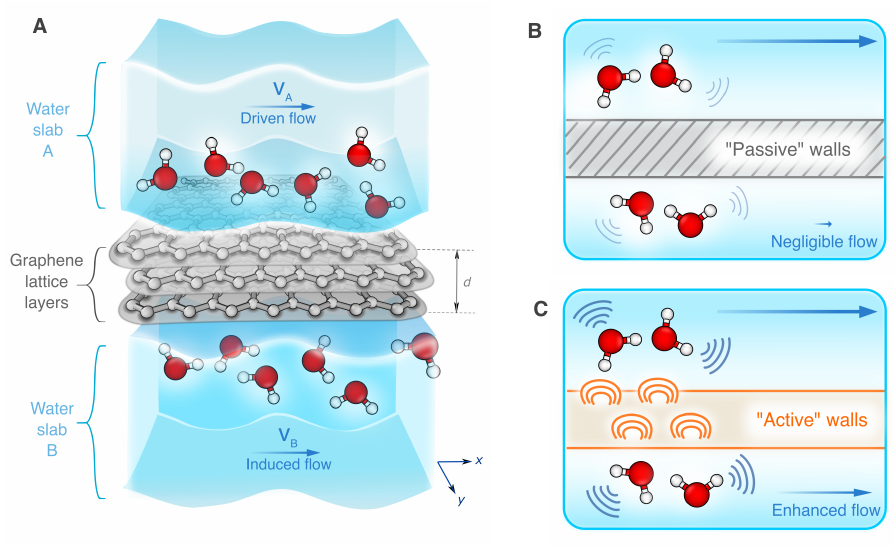} 
 \caption{\textbf{Principle of flow tunneling and role of the solid wall} 
 \tb{(A)}. Schematic of the system under study. Two water slabs $A$ and $B$, flowing at velocities $v_A$ and $v_B$, are separated by a solid wall of thickness $d$.  
 \tb{(B)} When the wall is ``passive'' and only acts as a static potential, 
 water on both sides of the solid can interact via fluctuating Coulomb forces. Direct momentum transfer between two slabs, however, 
 is negligible so the resulting flow tunneling effect is very small.
 \tb{(C)} When the wall is ``active'' through fluctuations in the solid
 coming from the electronic degrees of freedom (i.e. plasmons), the range and
 amplitude of flow tunneling can increase significantly 
due to the fluid--solid--fluid coupling. 
 }
 \label{main_model}
\end{figure*}

\vskip0.5cm
\noindent{\bf \large Flow tunneling through a passive wall} \\
The system we consider throughout is shown schematically in Fig. 1A, and comprises a liquid water slab on one side of $N$ two-dimensional solid layers (``fluid $A$''), with another slab of liquid water on the other side (``fluid $B$'').
Before investigating the role of interactions between the solid's and fluid's collective modes in mediating flow tunneling, we begin with a simpler question: To what extent does a driven flow in fluid $A$  directly induce a flow in fluid $B$? \tcb{To this end, we consider the case where } the solid layers have no internal degrees of freedom, and interact with the fluids only via a static potential. 
Taking the solid layers to lie in the $(x,y)$ plane, and assuming that fluids $A$ and $B$ flow with in-plane uniform velocities $\v_A$ and $\v_B$, respectively, 
we ask whether there is a net momentum transfer (or force) from fluid $A$ to fluid $B$. Such a force would originate from fluctuating Coulomb interactions between water slabs across the passive solid, and its computation in the framework of classical stochastic dynamics would be extremely involved \cite{Dean2010, Chen2015}. However, it can be readily estimated within an excitation perspective, using a quantum representation of the system.

The thermal charge fluctuations in each slab result in a fluctuating Coulomb potential acting on the other slab, that can be decomposed into evanescent plane waves of the form $\phi_{\mathbf{q}, \omega}(\mathbf{r},t) = \phi_0 e^{-qd} e^{i(\mathbf{q}\mathbf{r} - \omega t)}$, where $\mathbf{r}$ lies in the $(x,y)$ plane and $d$ is the separation between the outermost solid layers. The elementary bosonic excitations of these modes are the hydrons, in the same way that photons are elementary excitations of the electromagnetic field; $\phi_{\mathbf{q}, \omega}$ is then effectively the wavefunction of a hydron. \tcb{By analogy, since the hydron wavefunctions of fluids A and B overlap, water excitations can ``tunnel'' between the two
fluids.} The hydron transmission rate is then given by the canonical Landauer formula: 
\begin{equation}
 \Gamma^{A \to B} = \frac{1}{2 \pi \hbar} \sum_{\q} \int \d E \, (f^{A}_{\q}(E) - f^{B}_{\q}(E) ) \mathcal{T}_{\q} (E), 
 \label{landauer}
\end{equation}
where $f_{\q}(E)$ is the average number of hydrons of wavevector $\q$ and energy $E$ and $\mathcal{T}_{\q} (E) \propto e^{-2q d}$ is the dimensionless transmission coefficient, which (as a first approximation) scales as the squared overlap of the hydron wavefunctions. If $\v_A = \v_B = 0$, $f^{A}_{\q}(E) = f^{B}_{\q}(E) =  n_{\rm B}(\omega = E/\hbar)$, the Bose-Einstein distribution at temperature $T$, and there is no net hydron transmission from $A$ to $B$. Now, if $\v_A - \v_B = \Delta \v \neq 0$, the Bose-Einstein distributions experience a Doppler shift $\omega \mapsto \omega - \q \cdot \v_{A,B}$, so that  $\Gamma^{A \to B}$ is non-vanishing. This means that there is indeed a force, or net momentum transfer from $A$ to $B$, since a hydron of wavevector $\q$ carries a quantum of momentum $\hbar \q$.  To linear order in $\Delta \v$, and assuming a single energy scale $\hbar \omega_0 \ll k_B T$ for the hydrons, this force (per unit area) is given by
\begin{align}
 \frac{\tb{F}_{\rm hh}}{\mathcal{A}} &\approx \frac{\omega_0}{2 \pi} \frac{1}{\mathcal{A}} \sum_{\q} (\hbar \q) \left[ n_{\rm B}(\omega_0 - \q \cdot \Delta \v) - n_{\rm B}(\omega_0) \right] e^{-2qd} \nonumber \\
 &\approx \frac{3 k_{\rm B} T}{16 \pi^2 \omega_0 d^4} \Delta \tb{v}. 
 \label{Fhh_simple}
\end{align}
$F_{\rm hh}$ is the driving force for the flow tunneling effect: it induces the flow of fluid $B$ in response to the flow of fluid $A$. In the steady state, $F_{\rm hh}$ is balanced by the classical (roughness-induced) friction $F_{\rm cl} = - \lambda_{\rm cl} \mathcal{A} v_{\rm B}$ exerted on fluid $B$ by the solid wall, so that
\beq 
v_{\rm B}=\frac{\lambda_{\rm hh}}{\lambda_{\rm hh}+\lambda_{\rm cl}}v_{\rm A}, 
\label{vB_simple}
\eeq
where we have defined the hydron--hydron friction coefficient as $\lambda_{\rm hh} = F_{\rm hh} / (\mathcal{A} \Delta v)$. The quantum formalism has enabled us to obtain a first quantitative estimate for flow tunneling with minimal computations. The final result, however, describes a purely classical effect: Planck's constant is absent from Eq.~\eqref{Fhh_simple}. We may therefore assess the validity of our description using classical molecular dynamics (MD) simulations, where $\lambda_{\rm hh}$ is directly measured. As shown in the Supplementary Information (SI, Fig. S4), our prediction in Eq.~\eqref{Fhh_simple}  \tcb{matches} the simulation results at large separation $d$ between the slabs upon setting $\omega_0 = 0.3~\rm THz$, which is roughly the water Debye frequency. A more accurate analytical result that takes into account the full structure of the water fluctuation spectrum (SI Sec. IV) agrees with the simulation at arbitrary $d$.

Although qualitatively at odds with classical hydrodynamics, we find that, quantitatively, this form of flow tunneling through a passive solid is extremely short-ranged. \tcb{For example,} even assuming small roughness-based friction (e.g. $\lambda_{\rm cl}\approx 2.1\e{4}~\rm N\cdot s \cdot m^{-3}$ for graphene), we find that $v_{\rm B}$ is less than 1\% of $v_{\rm A}$ for $d\gtrsim 5$\,\AA, and thus negligible in all practical situations. We now show that the excitations of an ``active'' solid (Fig. 1C) drastically enhance the amplitude and range of flow tunneling, to the extent that it may become experimentally measurable.

\rev{The notion that the excitations of a solid wall can mediate momentum transfer between two liquids was in fact suggested more than 50 years ago by Andreev and Meierovich~\cite{Andreev1971}. They considered phononic excitations, whose ability to transfer momentum is limited by the acoustic impedance mismatch between the liquid and the solid: the predicted tunneling efficiency $v_{\rm B}/v_{\rm A}$ is at most $\sim 10^{-5}$, whatever the solid thickness (see SI, Sec. IV C). Here, we show that the physics are very different in the case of electronic excitations, leading to tunneling efficiencies up to $v_{\rm B}/v_{\rm A} \sim 1$. 
}

\begin{figure*}[t]
 \centering \includegraphics[width=\textwidth]{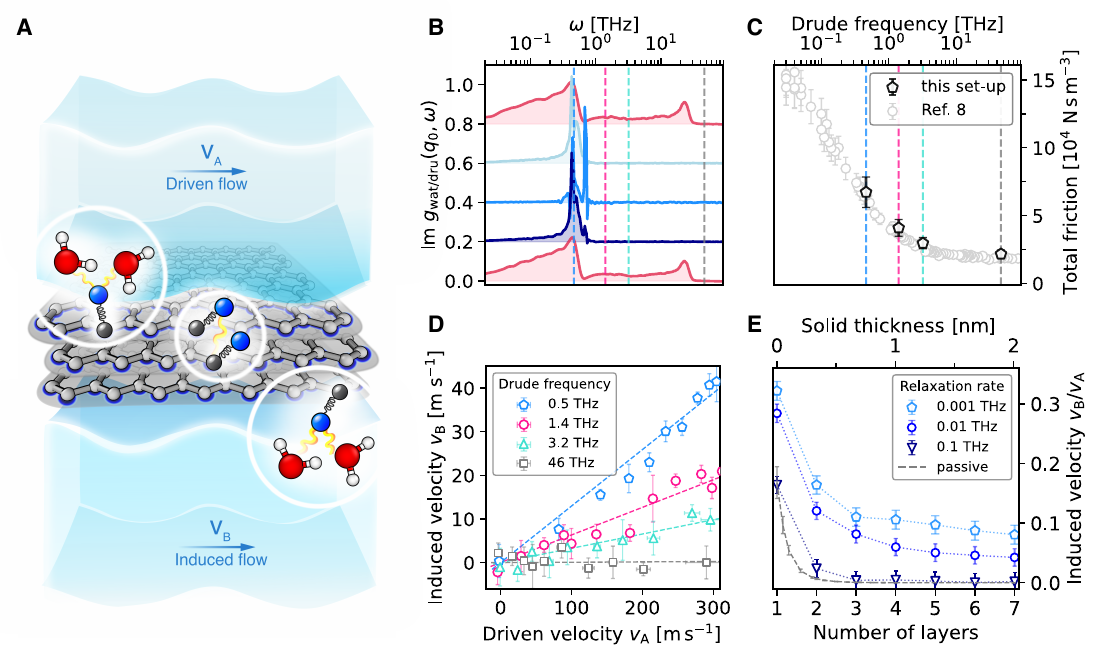} 
 \caption{\textbf{Molecular simulations of flow tunneling.} 
  \tb{(A)} Schematic representation of system simulated with Drude oscillators, used as a classical proxy for the quantum electron dynamics.  
 \tb{(B)} The surface excitation spectra as measured in 
 simulations of each component in a
 $N=3$ system: water slab A, each solid layer and water 
 slab B going from top to bottom. Here, the wavevector is $q_0 = 2.5$ nm$^{-1}$ and the Drude frequency is $\omega_{\rm p} = 0.5$ THz. The vertical lines indicate the other tested frequencies for the Drude oscillators. 
 \tb{(C)} The total solid--liquid friction corresponding to 
 the different Drude frequencies $\omega_p$ chosen for a system with $N=3$, 
 compared to result for a single water slab on a single graphene sheet in Ref.~\onlinecite{Bui2023}. The similarity of the results indicates that the friction is essentially determined by the interaction with the first solid layer.
  \tb{(D)} Induced velocity through $N=3$ layers of solid $v_B$ versus imposed velocity $v_A$ for the different Drude frequencies $\omega_p$, and fixed relaxation rate $\gamma= 10^{-3}$ THz. 
 \tb{(E)} Tunneling efficiency $v_B/v_A$ as a function of the number of layers $N$ for different relaxation rates $\gamma$ and fixed Drude frequency  $\omega_{\rm p} = 0.5$ THz (in blue) and through a passive solid (in gray).
 }
 \label{main_momentrumtransfer}
\end{figure*}

\vskip0.5cm
\noindent{\bf \large Molecular simulations of ``active'' flow tunneling} \\
Friction forces that arise from the dynamical coupling between the electronic excitations of the solid and \tcb{charge density fluctuations} in the liquid 
are nonadiabatic in nature. Such effects are beyond the Born-Oppenheimer approximation typically employed in molecular simulations, making the prospect of modeling these with explicit electronic dynamics, on time and length scales relevant to the problem at hand, a daunting task. Recently, however, a classical molecular dynamics scheme was shown to capture the most salient aspects of fluctuation--induced quantum friction \cite{Bui2023}. Here, we extend this approach to investigate ``active" flow tunneling, before providing a detailed, yet more general, theoretical account.

The solid wall is modeled as a stack of $N$ layers (with thickness 
$d = (N-1) d_0$ where $d_0$ is the spacing between two adjacent layers),
with each layer being composed of Lennard--Jones atoms 
arranged in a honeycomb lattice. 
Electron dynamics are mimicked by giving each 
atom a positive charge, and attaching to it a fictitious Drude
particle of equal and opposite charge via a harmonic spring (Fig. 2A) \cite{Lamoureux2003,Misra2017}.
Relaxation processes in the solid (electron-phonon and impurity scattering,
umklapp processes \cite{kittel2004introduction}) are taken into 
account implicitly through an effective damping rate $\gamma$ for 
the Drude oscillators. This gives the solid prototypical charge 
fluctuations described by a single plasmon-like mode at a frequency
$\omega_{\rm p}$, which can be adjusted by tuning the mass of the Drude particle. \rev{
While the Drude model is a crude representation for a realistic plasmon and its dispersion behavior,  it allows us to capture the essential physics since its principal mode can be tuned to overlap in frequency with the water's surface response}
(Fig. 2B), thereby controlling the degree of dynamical coupling between the solid and the liquid. As seen in Fig. 2C, as $\omega_{\rm p}$ approaches the THz regime from above, the total solid--liquid friction increases from its ``classical'' surface-roughness value $\lambda_{\rm cl}$: this \tcb{extra contribution} is the fluctuation-induced (quantum) friction $\lambda_{\rm qf}$~\cite{Bui2023,Kavokine2022}. Momentum transfer from the liquid to the solid is therefore enhanced by matching the frequencies of their respective charge fluctuations.

We then performed non-equilibrium molecular dynamics simulations with a pressure gradient applied to fluid $A$, and measured the resulting non-equilibrium steady state flow velocities in both fluids $A$ and $B$. In the absence of Drude oscillators, no induced flow in fluid $B$ could be measured for a solid thicker than a single layer, in line with our prediction in Eqs.~\eqref{Fhh_simple} and \eqref{vB_simple}. However, when the Drude frequency was set in the range of water's Debye modes, we observed a large induced flow even through much thicker solids, up to $N = 7$ layers (Fig. 2D). For example, $v_B \approx 0.1 \,v_A $ for a $2\,$nm thick ($N=7$) solid with $\omega_{\rm p} = 0.1\,\rm THz$ and relaxation rate $\gamma = 10^{-3}\,\rm THz$. The induced velocity $v_{B}$ scaled linearly with the driven velocity $v_{\rm A}$ in the investigated range (Fig. 2D). 

Our simulations thus reveal that the coupling to the solid's charge fluctuation modes does not simply take momentum away from fluid $A$ through friction. Momentum is in fact accumulated in those modes, and part of it is transmitted to fluid $B$, resulting in flow tunneling. The amount of momentum accumulation is sensitive to the relaxation rate $\gamma$, with faster relaxation leading to weaker flow tunneling (Fig. 2E). \tcb{While in our simulations, the Drude particles themselves do not flow, in a real solid, there would be propagation of both collective plasmons and single electrons. The latter would induce an electric current parallel to the surface,} akin to the Coulomb drag phenomenon~\cite{Narozhny2016}. 
\rev{We finally note that the momentum transfer is measured to be important in spite the fluid flow being transverse to the direction of momentum transfer across the layers, a point further confirmed by the theoretical modelling. }

\begin{figure*}[t]
 \centering 
 \includegraphics[width=\textwidth]{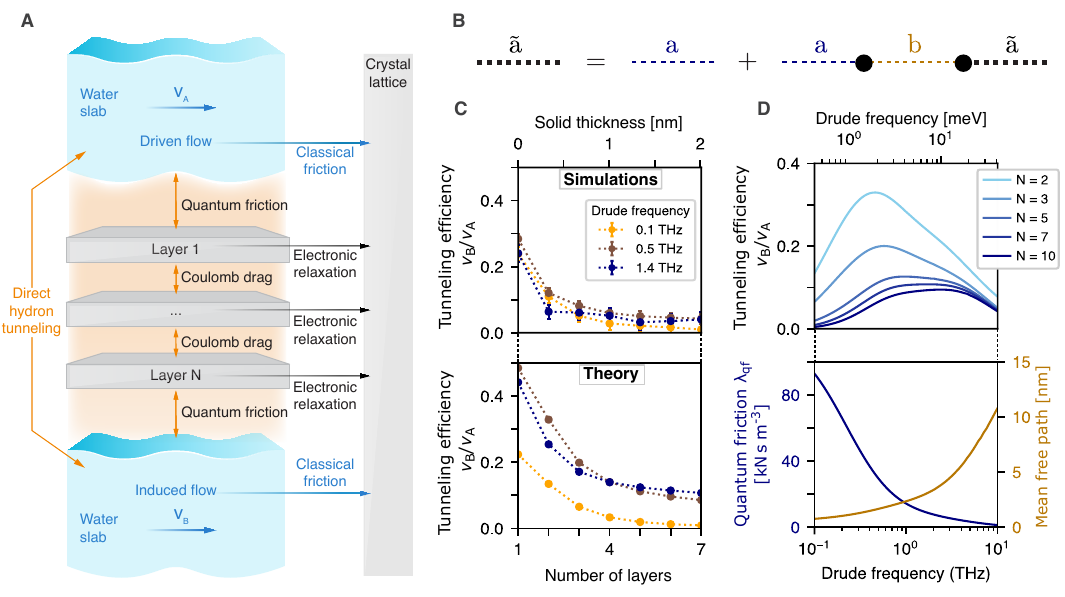} 
 \caption{\textbf{Quantum theory of flow tunneling.} 
 \tb{(A)} Diagram representing the momentum fluxes in the liquid--solid--liquid system. Momentum can be transferred from fluid $A$ to fluid $B$ sequentially through the $N$ solid layers, or through direct hydron tunneling if the solid is thin. The processes that dissipate momentum are represented by the horizontal arrows. 
 \tb{(B)} Dyson equation for the surface response function of layer $a$ renormalized by the interactions with layer $b$. A thin line represents a bare response function (in blue for $a$ and gold for $b$) while the black thick lines represent the renormalized response function of $a$. 
   \tb{(C)} Tunneling efficiency $v_B/v_A$ as a function of the number of layers for different Drude frequencies $\omega_p$ and a relaxation rate of $\gamma=10^{-2}$ THz. Top: simulation data. Bottom: theoretical prediction. 
     \tb{(D)} Tunneling efficiency $v_B/v_A$ (top) as a function of the Drude frequency $\omega_p$ for a relaxation rate of $\gamma=10^{-2}$ THz. There is an optimal Drude frequency because of the trade-off between the two quantities plotted at the bottom: the quantum friction coefficient (in blue) and the hydron mean free path inside the solid (in gold).
 }
 \label{main_theory}
\end{figure*}

\vskip0.5cm
\noindent{\bf \large  Many-body quantum theory of flow tunneling} \\
Guided by the simulation results, we now develop a theory of flow tunneling through an active solid. \rev{Within our formally quantum picture, flow tunneling through a passive solid amounted to coherent hydron transport between the two fluids -- it was therefore described by a Landauer formula. The active solid now plays the role of a "junction" placed between the two hydron reservoirs (the fluids). As highlighted by the strong dependence of the simulation results on the relaxation parameter $\gamma$, the hydron transport through this junction cannot be considered coherent, and its description therefore requires going beyond the Landauer formalism. In order to account for decoherence, we model the solid by a layered structure in our simulations. The hydrons are transported coherently between the layers, but they can undergo inelastic scattering (i.e., decoherence) within each of the layers.} Technically, we use the Keldysh formalism of perturbation theory, which has proven to be an asset in the study of non-equilibrium solid-liquid systems~\cite{Kavokine2022,Coquinot2023}. Our computation is fully detailed in the SI, Sec. III-V; here, we outline the main steps.

The system is described in terms of the fluctuating charge densities $n(\rr,t)$ of both the liquid slabs and the solid layers. It is governed by the Hamiltonian comprising all Coulomb interactions: between the water and the solid, between the two water slabs and between the different solid layers. To keep the computations tractable, we assume that a solid layer interacts only with its nearest neighbors (Fig. 3A). Our goal is to evaluate the average Coulomb force exerted by the $N^{\rm th}$ solid layer on the fluid slab $B$, in the non-equilibrium state where fluid $A$ flows at velocity $v_A$: 
\beq
\tb{F}_{NB} = \int \d \rr_N \d \rr_B \langle n_N(\rr_N, t) V(\rr_N - \rr_B) n(\rr_B, t)\rangle, 
\label{basic_force}
\eeq
where $\rr_N$ and $\rr_B$ are the spatial coordinates in layer $N$ and fluid slab $B$, respectively, and $V$ is the Coulomb potential. To this end, we formally quantize the charge densities as free Gaussian fields -- this is an approximation that amounts to neglecting interactions between excitations. \rev{The liquid and the solid are then fully characterized by their charge density correlation functions, which can be evaluated starting from the microscopic model of one's choice. In the following, we will describe the solid by the correlation function of the Drude oscillator model, to allow for direct comparison with the simulations. However, one could model each solid layer as a 2D electron gas with appropriate electron-electron and electron-phonon interactions (accounting for decoherence), so as to describe realistic solid-state systems that cannot be treated classically, such as few-layer transition metal dichalcogenides or MXenes.} 

 The model defined in this way is in fact integrable (the Hamiltonian is quadratic in the charge densities), so that the non-equilibrium state of the system can be determined exactly using the Keldysh formalism. 
In practice, we perturbatively expand the average in Eq.~\eqref{basic_force} in powers of the Coulomb interaction, and exactly resum the infinitely many terms of this expansion. The basic building blocks of the expansion are the {\it surface response functions} $g$ of each of the water slabs and solid layers. These are appropriately normalized charge density correlation functions: $g(\q,\omega) \sim V_q \langle n_{\q} n_{-\q} \rangle_{\omega}$, where $V_q = e^2/(2\epsilon_0 q)$ is the Fourier-transformed Coulomb potential and the $n_{\q}$ are charge density operators. Precise definitions of these quantities are given in the SI, Sec. III.C. In the Keldysh formalism, these correlation functions possess three components ($g^R, g^A, g^K$), corresponding to different time orderings of the operators. 
In the non-interacting equilibrium state, the Drude response function of a solid layer is 
\beq
g^{R}_{\rm eq, ~ Drude} = \frac{\omega_{\rm p}^2f(q)}{\omega_{\rm p}^2 - \omega^2 - 2 i \gamma \omega}, 
\eeq
and the water surface response function is modeled, following Ref.~\onlinecite{Kavokine2022}, as a sum of two Debye peaks: 
\beq
g^{R}_{\rm eq, ~ Water} = \sum_{i = 1,2} \frac{f_i(q)}{1 - i \omega/\omega_{\mathrm{D}, i}}. 
\eeq
The values of the parameters and expressions of the functions $f(q)$ are given in the SI, Sec. IV. 
The imposed flow in the fluid $A$ 
and the induced flow in the fluid $B$ 
are described, as before, by a Doppler shift in the non-equilibrium water surface response function versus its equilibrium expression: $g(\q,\omega) = g_{\rm eq}(\q,\omega - \q \tcr{\cdot}\v_{A,B})$. The expansion proceeds in two steps. \rev{First, the non-equilibrium response functions of the $N$ solid layers are determined by solving a series of Dyson equations, accounting for the inter-layer coupling via the electrostatic interactions and decoherence process throughout the solid}. These equations have a very simple diagrammatic representation (Fig. 3B) -- since the Hamiltonian is quadratic in the density operators -- and they are algebraic in Fourier space. 
\rev{For example, the renormalisation of the layer $i + 1$ by the layer $i$ for the retarded component is given by $g^R_{i+1} = g^{R,\rm eq}_{i + 1} + g^{R,\rm eq}_{i + 1} g^R_i g^R_{i+1}$, and the expression for the Keldysh component is given in the SI, Sec. III D. }
Then, the expansion of Eq.~\eqref{basic_force} is carried out in terms of those response functions, yielding the force acting on the B fluid across the $N$ layers of material:
\beqa 
\frac{\tb{F}_{NB}}{\A}&=&\int\frac{\dd\q}{(2\pi)^2} (\hbar\q)  (\Gamma_{NB}(\q) - \Gamma_{BN}(\q))
\label{generic_friction}
\eeqa
with
\beqa 
 \Gamma_{\rm ab}(\q)=\int\frac{\dd\omega}{4i\pi} \frac{\im{g_{\rm a}^{\rm R}(\q,\omega)}g^{\rm K}_{\rm b}(\q,\omega)}{|1-g_{\rm a}^{\rm R}(\q,\omega)g_{\rm b}^{\rm R}(\q,\omega)|^2}.
 \label{gamma_ab}
\eeqa
Eqs.~\eqref{generic_friction} and \eqref{gamma_ab} are our main theoretical result, which has general validity beyond the particular Drude model of the solid we have considered so far. It is a far-from-equilibrium generalization of quantum friction~\cite{Kavokine2022}, which echoes the Landauer formula in Eq.~\eqref{landauer}. For true interacting electrons, our result is valid at the level of a self-consistent Hartree approximation. $\tb{F}_{NB}$ depends on both $v_{\rm A}$ and $v_{\rm B}$ and can be expanded to linear order in these velocities, defining two friction coefficients: $\tb{F}_{NB}/\A=\lambda_{\rm drive}\v_{\rm A}- \lambda_{\rm qf}\v_{\rm B}$. The coefficient
$\lambda_{\rm qf}$ is the fluctuation-induced (quantum) friction coefficient between the fluid slab $B$ and the $N^{\rm th}$ solid layer at equilibrium. \rev{The coefficient $\lambda_{\rm drive}$ accounts for the "remote drag" exerted by fluid $A$ on fluid $B$ -- its expression is cumbersome (see SI Sec. V), but we provide in the following a scaling estimate that allows us to draw practical conclusions.} 

Physically, the solid gives momentum to the liquid, but then takes some of it back through quantum friction. Fluid $B$ is also subject to an additional force due to direct hydron tunneling ($\lambda_{\rm hh}$) if the solid is thin, and to the classical roughness-based friction ($\lambda_{\rm cl}$) on the $N^{\rm th}$ layer. Momentum conservation then imposes 
\beq
\v_{\rm B}=\frac{\lambda_{\rm drive}+\lambda_{\rm hh}}{\lambda_{\rm qf}+\lambda_{\rm cl}+\lambda_{\rm hh}}
\v_{\rm A}. 
\label{vBA}
\eeq
Eq.~\eqref{vBA} is our theoretical prediction for the flow tunneling effect. In Fig. 3C, we compare the theoretical predictions against simulation results. Given the simplifying assumptions in the theoretical model (i.e., nearest-neighbor interactions between graphene layers, and a harmonic approximation for water's dielectric fluctuations) the agreement is remarkable, and suggests Eq.~\eqref{vBA} captures the essential physics of the flow tunneling effect.

\vskip0.5cm
\noindent{\bf \large Conditions for optimal \tcb{flow tunneling}} \\
Having established a theoretical framework, we may now assess
the precise role of the solid's electronic properties in 
determining the range of flow tunneling. 
Fig. 3D shows the prediction for the tunneling efficiency $v_B/v_A$ as a function 
of the Drude plasmon frequency $\omega_{\rm p}$, at fixed
$\gamma = 10^{-2}$ THz. 
Interestingly, it exhibits a maximum at  $\omega_{\rm p} \sim 0.3~\rm THz$ where
the Drude frequency is in resonance with
the water Debye frequency (see Fig. 2B).
 Moreover, we observe that at small $\omega_{\rm p}$ the flow tunneling efficiency decreases with increasing number 
 of solid layers $N$, while at large $\omega_{\rm p}$ it is nearly independent of $N$ for $N \leq 7$. 

These features can be understood if flow tunneling is represented as the transmission of discrete excitations -- hydrons -- between fluid $A$ and fluid $B$. In Eq.~\eqref{vBA} the friction coefficient $\lambda_{\rm drive}$ is effectively the rate at which the solid injects momentum-carrying hydrons into fluid $B$. Hydrons are injected into the solid from fluid $A$ with a rate $\lambda_{\rm qf}$. However, they only reach fluid $B$ if they are not scattered during their residence time $\tau$ inside the solid, which happens at a rate $\gamma$. If the solid is moderately \tcb{thin}, hydrons reach the opposite boundary quickly, and $\tau$ is the time it takes for a hydron to exit into fluid $B$: $\tau \approx \tau_{\rm sl} \sim \lambda_{\rm qf}^{-1}$. For thicker solids, most hydrons get scattered on their way from $A$ to $B$ so that $\tau \propto N$, and we may define a mean free path $\ell$ according to $\gamma \tau \equiv d /\ell$. Altogether (and assuming $N > 1$ so that $\lambda_{\rm hh}$ can be neglected) we obtain an asymptotic expression for the tunneling efficiency, valid for thick solids:
\beq
\frac{v_B}{v_A} \approx \frac{\lambda_{\rm qf}}{\lambda_{\rm qf} + \lambda_{\rm cl}} e^{-(\gamma \tau_{\rm sl} + d/\ell) },
\label{vBA_simple}
\eeq
with $\tau_{\rm sl}$ and $\ell$ both depending on the Drude frequency $\omega_{\rm p}$ and relaxation rate $\gamma$. 
For thin solids (small $N$, as in our simulations), we obtain a more accurate scaling expression that accounts for the discreteness of the layers (SI Sec. V. D), but the data is still well described by Eq.~\eqref{vBA_simple} (Fig. S6F). Within the Drude model of the solid, $\lambda_{\rm qf}$ is a strictly decreasing function of $\omega_{\rm p}$ (Fig. 3D): hydron injection into the solid is most efficient for lower Drude frequencies. \rev{However, the hydron mean free path is shorter for lower $\omega_{\rm p}$ (Fig. 3D): lower frequency modes take longer to transmit their excitations to the next layer. The thicker the solid, the more flow tunneling is favored by higher plasmon frequencies.} This trade-off between hydron injection rate and mean free path accounts for the 'resonant', bell-shaped dependence of the tunneling efficiency on $\omega_{\rm p}$. 

Going further, the result in Eq.~\eqref{vBA_simple} allows us to draw general conclusions regarding the range of flow tunneling. This range is limited by the scattering rate $\gamma$ of excitations inside the solid: flow can in principle tunnel through an arbitrarily thick solid if there is no dissipation inside. This is never the case in practice, and no tunneling is observed if the solid is thicker than a few times the mean free path $\ell$. For $d \gtrsim \ell$, the tunneling efficiency decreases exponentially with $d$ (see Fig. 3C at low $\omega_{\rm p}$): the hydron transport is diffusive. However, the mean free path can sometimes become very large (Fig. 3D at large $\omega_{\rm p}$). Then, it is realistic to have $d \ll \ell$, and the hydron transport becomes ballistic. The tunneling efficiency is then limited by the solid-liquid crossing time $\tau_{\rm sl} \sim \lambda_{\rm qf}^{-1}$: hydrons are scattered as they wait in the solid to cross into fluid $B$. $v_B/v_A$ is then independent of $d$ as seen in Fig. 3D at high Drude frequencies. \rev{But in both the diffusive and the ballistic regimes, the tunneling efficiency is ultimately determined by comparing the residence time of an excitation inside the solid to its inelastic scattering (or dephasing) rate.}

We have thus identified the three qualitative determinants of flow tunneling: hydron injection rate, hydron mean free path $\ell$, and hydron exit rate $\tau_{\rm sl}^{-1}$, which combine to predict the tunneling efficiency in Eq.~\eqref{vBA_simple}. We may now discuss the possibility of flow tunneling in a realistic solid-liquid system -- such as two slabs of water separated by a graphene multilayer -- based on these three ingredients. Within our model, the hydron injection rate is set by the quantum friction coefficient $\lambda_{\rm qf}$. However, this ignores electron-phonon coupling in the solid and the resulting phonon drag effect~\cite{Coquinot2023}, which effectively replaces $\lambda_{\rm qf}$ by $\sim \lambda_{\rm cl}$ in Eq.~\eqref{vBA_simple}. Because the electron-phonon and electron-hydron scattering rates typically have similar values~\cite{Coquinot2023}, we further expect $\gamma \tau_{\rm sl} \sim 1$. Therefore, the one key determinant of the tunneling efficiency is in fact the hydron mean free path, for which our theory may provide a quantitative estimate (SI Sec. V.D): in the Drude model framework, we find the phenomenological scaling $\ell \approx \ell_0 \sqrt{\omega_p/\gamma}$, with $\ell_0 = 0.26~\rm nm$. Ignoring dispersion effects, the graphene plasmon mode may be approximately described by $\omega_{\rm p} = 100~\rm THz$ and $\gamma = 0.6~\rm THz$~\cite{Ni2018}, yielding $\ell \approx 4~\rm nm$. We note that this is likely an underestimation, as it ignores electron transport perpendicular to the graphene layers. Overall, we may expect non-negligible flow tunneling through a 10 nm thick graphene wall, which can be readily obtained in nanofluidic systems using, e.g., van der Waals assembly.

\vskip0.5cm
\noindent{\bf \large Conclusions} \\
Using a combination of many-body quantum theory and molecular simulations, we have shown that the flow of one liquid can induce the flow of another liquid through a solid wall of nanoscale thickness -- a phenomenon termed 'flow tunneling'. Classical hydrodynamics have so far been found to hold surprisingly well down to $1\,$nm wide channels. Our prediction implies that, in systems of multiple channels, the classical framework of hydrodynamics may qualitatively break down if the walls separating the channels are thinner than $\sim 10~\rm nm$. The physical origin of this breakdown lies in the coupling of the liquid charge fluctuations to the solid wall's electronic excitations. 
%
Beyond the fundamental importance of flow tunneling as an effect beyond hydrodynamics, this property
is expected to be at play in global transport across nanoscale fluidic networks, in particular across membranes made of lamellar materials such as graphene oxides or MXenes \cite{Ouyang2021, Gogotsi2019}. 
Flow tunneling is also a new and promising lever for manipulating liquids at the nanoscale via their dielectric spectrum, and not based on the nature and characteristics of individual molecules.


\vskip0.5cm
\noindent{\bf \large Acknowledgements} \\
\small
The authors thank Mischa Bonn for many fruitful discussions. 
We thank Lucy Reading-Ikkanda for help with figure preparation. 
We are also grateful for computational support 
from the UK national high performance computing service, ARCHER2, for which access was obtained via the UKCP consortium and funded by
EPSRC grant ref EP/X035891/1.
\textbf{Funding:}
B.C., D.T., A.M. and L.B. acknowledge support from ERC project {\it n-AQUA}, grant agreement $101071937$.
B.C. acknowledge support from the CFM foundation. A.T.B. acknowledges
funding from the Oppenheimer Fund and Peterhouse
College, University of Cambridge.  
The Flatiron Institute is a division of the Simons Foundation. 
S.J.C. is a Royal Society
University Research Fellow (Grant No. URF\textbackslash
R1\textbackslash 211144) at the University of Cambridge. 
\textbf{Data availability:}
The simulation data that support the findings of this study are openly available at the University of Cambridge Data Repository at https://doi.org/10.17863/CAM.113204.
\textbf{Author contributions:} A.M., N.K., S.J.C, and L.B. conceptualized the project; B.C. carried out the theoretical analysis; A.T.B. designed and performed the molecular simulations; D.T. performed supporting analyses; B.C., A.T.B, A.M., N.K., S.J.C, and L.B. wrote the paper.
\textbf{Competing interests:}
 The authors declare that they have no competing interests.

\normalsize
\bibliography{main_bibliography.bib}

\end{document}


\title{Supplementary Information\\
Momentum tunneling between nanoscale liquid flows}

\author{Baptiste Coquinot$^{1,2\dagger}$, Anna T. Bui$^{3,\dagger}$, Damien Toquer$^{1}$,  Angelos Michaelides$^3$, Nikita Kavokine$^{2,4*}$, Stephen J. Cox$^{3*}$ and Lyd\'eric Bocquet$^1$}
\email{sjc236@cam.ac.uk, nikita.kavokine@mpip-mainz.mpg.de, lyderic.bocquet@ens.fr}
\affiliation{$^1$Laboratoire de Physique de l'Ecole Normale Sup\'erieure, 24 rue Lhomond, 75005, Paris, France}
\affiliation{$^2$Max Planck Institute for Polymer Research, Ackermannweg 10, Mainz, Germany}
\affiliation{$^3$ Yusuf Hamied Department of Chemistry, University of Cambridge, Cambridge CB21EW, United Kingdom}
\affiliation{$^4$Center for Computational Quantum Physics, Flatiron Institute, 162 5$^{\rm th}$ Avenue, New York, NY 10010, USA\\
$\dagger$: these authors contributed equally}

\maketitle

\tableofcontents


\section{Simulations with passive walls}

\subsection{Set-up}

\begin{figure*}[t]
 \centering \includegraphics{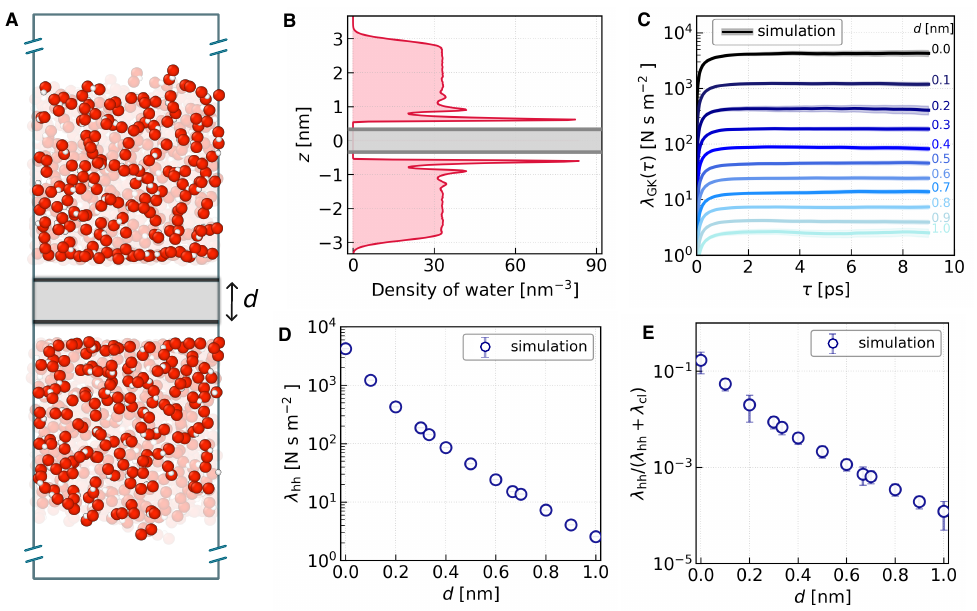} 
 \caption{\textbf{Simulations with passive walls.}
 \tb{(A)} The view from the side of the simulation set-up.  
 Oxygen and hydrogen atoms are in red and white respectively.
 \tb{(B)} The planar density profile of the water slabs.
 \tb{(C)} The Green--Kubo friction integrals 
 $\lambda_{\rm GK}(\tau)$ for different values of thickness $d$.
 \tb{(D)} The hydron--hydron friction coefficient
 $\lambda_{\rm hh}$ as a function of $d$.
 \tb{(E)} The flow tunneling efficiency 
 when the walls are passive calculated from Eq.~5.}
 \label{SI_fig1}
\end{figure*}

For simulations with passive solid walls, we consider the 
system in Fig.~1A with two water slabs, each with 517 water 
molecules (thickness $\sim2.5\,\mrm{nm}$) separated by a
passive solid of thickness $d$ in the $z$ direction. The top 
solid surface interacts with
water slab A through a Lennard--Jones 9-3 potential
\begin{equation}
    u_{\rm hi}(z) = \epsilon_{\rm wall} \left[ \frac{2}{15} \left(\frac{\sigma_{\rm wall}}{|z-z_{\rm hi}|} \right)^{9}-\left(\frac{\sigma_{\rm wall}}{|z-z_{\rm hi}|} \right)^{3}\right] \quad \quad z>z_{\rm hi},
\end{equation}
and the bottom surface with water slab B through
\begin{equation}
    u_{\rm lo}(z) = \epsilon_{\rm wall} \left[ \frac{2}{15} \left(\frac{\sigma_{\rm wall}}{|z-z_{\rm lo}|} \right)^{9}-\left(\frac{\sigma_{\rm wall}}{|z-z_{\rm lo}|} \right)^{3}\right] \quad \quad z<z_{\rm lo},
\end{equation}
where $z_{\mrm{hi}}=z_{\mrm{lo}}+d$, $\epsilon_{\mrm{wall}}$
is the interaction strength and $\sigma_{\mrm{wall}}$ is
the interaction range. We choose 
$\epsilon_{\mrm{wall}}=1.3\,\mrm{kcal\,mol}^{-1}$ and 
$\sigma_{\mrm{wall}}=3.19\,\mrm{\AA}$, which gave
a similar water density profile at the liquid--solid
interface (Fig.~1B) as the water--carbon interface
of Ref.~\cite{Bui2023}.
However, the conclusion does not change when
these parameters deviate from the quoted values. 

\subsection{Simulation details}

Simulations were carried out with the \texttt{LAMMPS}
simulations package \cite{Thompson2022}. The orthorhombic cell
has dimension 
$\approx26\times25\times\SI{300}{\angstrom\cubed}$.
Water--water interactions were described with the SPC/E 
water model \cite{Berendsen1987}. 
The geometry of water molecules was constrained using 
the \texttt{RATTLE} algorithm \cite{Andersen1983}. 
All Lennard-Jones interactions were truncated and shifted at 
\SI{10}{\angstrom}. 
Electrostatic interactions were cut off at \SI{10}{\angstrom} and 
long-ranged interactions were evaluated using particle--particle 
particle--mesh Ewald summation \cite{hockney1988} such that
the RMS error in the forces was a factor of $10^5$ smaller 
than the force between two unit charges separated by a 
distance of \SI{1.0}{\angstrom} \cite{Kolafa1992}.
The Yeh--Berkowitz correction \cite{Yeh-Berkowitz1999} 
was used to decouple the electrostatic interactions
between the whole system and its periodic images.

The simulations were carried out in the canonical 
(NVT) ensemble for different $d$ values, 
where the temperature was held at $\SI{300}{K}$. 
Two separate Nos\'{e}--Hoover thermostats
\cite{Shinoda2004,Tuckerman2006} were applied to each water
slab. Each thermostat is a Nos\'{e}--Hoover chain with 10 
thermostats and a damping constant of $\SI{0.1}{\pico\second}$.
Dynamics were propagated using the velocity Verlet algorithm with a 
time-step of $\SI{1}{\femto\second}$. 
Each system was equilibrated for $\SI{100}{\pico\second}$ 
and the subsequent $\SI{10}{\nano\second}$ was used 
for analysis to give the results presented.

\subsection{Computation of the hydron--hydron friction coefficient}

For each equilibrium MD simulation, the friction coefficient was
evaluated through the Green--Kubo formula \cite{Bocquet1996}
involving the time integral of the force autocorrelation function
\begin{equation}
  \label{Eq:lambdaGK}
  \lambda_{\mrm{GK}}(\tau) = \frac{1}{\mcl{A}k_{\rm B}T}
  \int_0^\tau\!\mrm{d}t\,\langle F(0)\cdot F(t)\rangle,
\end{equation}
where $\mathcal{A}$ is the interfacial lateral area,
$\langle\cdots\rangle$ indicates an ensemble average
and $F(t)$ denotes the instantaneous lateral 
force exerted on the liquid by the solid at time $t$.
$F(t)$ is evaluated as the total
summed force acting on all water molecules of a given 
configuration averaged over both in-plane dimensions $(x,y)$
and is saved at every time-step ($\SI{1}{\femto\second}$).
In principle, the hydron--hydron friction coefficient 
is recovered at the long-time limit
\begin{equation} 
\lambda_{\rm hh} = \lim_{\tau\to\infty} \lambda_{\mrm{GK}}(\tau).
\end{equation}
However, at long times, the integral in Eq.~\ref{Eq:lambdaGK} decays
to zero due to the finite lateral extent of the system 
\cite{Bocquet1997,Espanol2019} so we evaluated  
$\lambda_{\mrm{GK}}(\tau)$ to a plateau as is commonly done
\cite{Falk2012,Tocci2014,Poggioli2021} (see Fig.~1C).
The error bars correspond to the statistical errors obtained
from splitting the entire trajectory into 100 blocks
such that each block is $100\,\mrm{ps}$ long.

The hydron--hydron friction coefficient $\lambda_{\mrm{hh}}$
from the simulations is shown as a function of the solid 
thickness $d$ in Fig.~1D. By assuming a surface roughness-based
classical friction of $\lambda_{\mrm{cl}}\approx2.2\times10^4\,\mrm{N\,s\,m^{-3}}$
(justified in a later section with simulations with explicit carbon atoms)
and using Eq.~3 in the main text
\begin{equation}
    v_{\rm B}=\frac{\lambda_{\rm hh}}{\lambda_{\rm hh}+\lambda_{\rm cl}}v_{\rm A},
\end{equation}
we show the prediction from simulations for
the flow tunneling efficiency $v_{\rm B}/v_{\rm A}$ in Fig.~1E.
The scaling relation between $v_{\rm B}/v_{\rm A}$  and $d$
will be compared with the theory in a later section.

\section{Simulations of flow tunneling}

\begin{figure*}[b]
 \centering \includegraphics{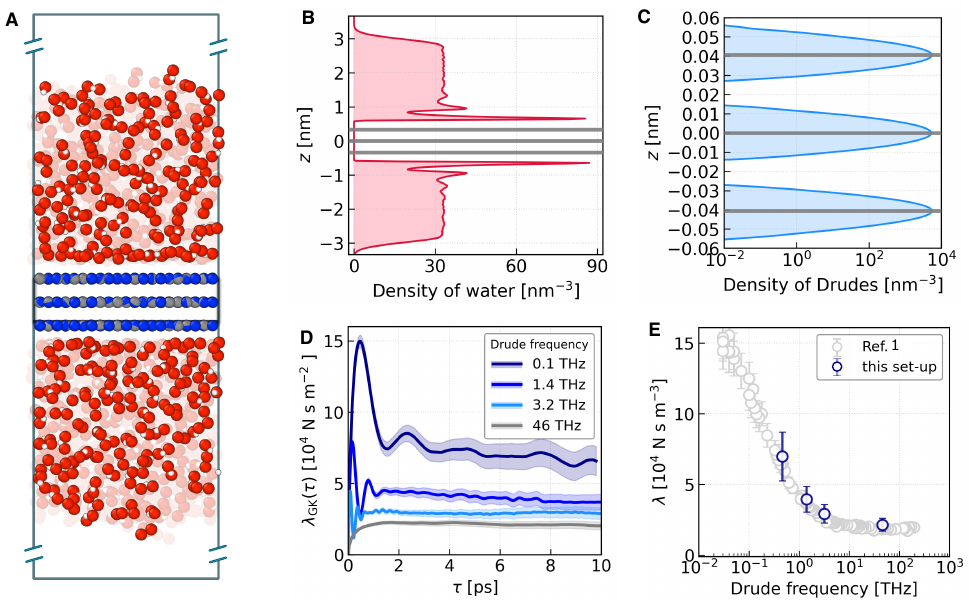} 
 \caption{\textbf{Simulations with active walls
 with three layers of graphene.}
 \tb{(A)} The view from the side of the simulation set-up.  
 Oxygen, hydrogen, carbon atoms and Drude particles are in red,
 white, grey and blue, respectively
 \tb{(B)} The planar density profile of the water slabs (grey lines indicate location of the graphene sheets).
\tb{(C)} The planar density of the Drude particles about each graphene
layer. The origins for the top and bottom layers are shifted to 
show both three distributions on the same plot.
 \tb{(D)} The Green--Kubo friction integrals 
 $\lambda_{\rm GK}(\tau)$ for different values of Drude
 frequecies $\omega_{\mrm{p}}$.
 \tb{(E)} The liquid--solid friction coefficient on 
 water slab A,  $\lambda$, as a function of the Drude
 frequency $\omega_{\rm p}$. Results for friction of one 
 water slab on a single graphene layer from Ref.~\onlinecite{Bui2023}
 is also shown.}
 \label{SI_fig2}
\end{figure*}

\subsection{Set-up}

To explore the flow-tunneling effect when the solid's charge density
fluctuations can play an active role, we consider a system of $N$ solid layers sandwiched between two films of 
water as shown in Fig.~2A. The layers are modeled with the lattice and Lennard-Jones parameters of graphene. 
Water is modelled with the SPC/E model \cite{Berendsen1987} and the
water--carbon interaction is described with a 12-6 Lennard-Jones
potential with the Werder parameters \cite{Werder2003}.
The dielectric response of the solid is described
with the classical Drude model \cite{Lamoureux2003}: each C atom carries 
a charge $+Q_{\rm D}$ and is attached to a Drude particle of 
mass $m_{\rm D}$ and charge $-Q_{\rm D}$ with a harmonic spring
with force constant $k_{\rm D}$.
Following Ref.~\onlinecite{Bui2023},
we set $Q_{\rm D} = 1.852\,e$ 
and $k_{\rm D} = \SI{4184}{\kJmol}\mrm{\AA}^{-2}$, 
which have been shown to recover the polarizability tensor of a 
periodic graphene lattice \cite{Misra2017}. The resulting 
water density profile and 
distribution of Drude particles about their core carbon atoms 
are shown in Fig.~2B and Fig.~2C respectively. As done in
Ref.~\onlinecite{Bui2023}, we treat $m_{\rm D}$ as a 
free parameter that tunes the frequency 
of each individual oscillator, which in turn controls the
frequency of the single plasmon-like mode $\omega_{\rm p}$.
The masses we have used and their resulting values for
 $\omega_{\rm p}$ are given in Table~1.

\begin{table}[H]
    \centering
    \begin{tabular}{c c}
    \hline
    \hline
    Mass $m_{\mrm{D}}$ [amu] & Drude frequency $\omega_{\rm p}$ [THz]  \\
    \hline
    5      &   46       \\
    1000   &   3.2      \\
    5000   &   1.4      \\
    50000  &   0.5      \\
    500000 &   0.1      \\
    \hline
    \hline
    \end{tabular}
\caption
{\textbf{Parameters for masses in simulations.}
The frequencies are extracted from the location of the 
principal peak in the solid layers' surface response functions.}
\end{table}

\subsection{Equilibrium simulation details}

Simulations were carried out with the \texttt{LAMMPS}
simulations package \cite{Thompson2022}. 
The orthorhombic cell
has dimension 
$\approx26\times25\times\SI{300}{\angstrom\cubed}$.
Water--water interactions were described with the SPC/E 
water model \cite{Berendsen1987}. The geometry of water 
molecules was constrained using the \texttt{RATTLE} algorithm
\cite{Andersen1983}. The carbon positions of the sheet were
fixed. Water--carbon interaction was modeled with Werder parameters 
\cite{Werder2003}. 
The classical Drude oscillator model \cite{Lamoureux2003, Dequidt2016}
was employed for the polarisability of the carbon
with Thole damping \cite{Thole1981} using parameters 
from Misra and Blankschtein \cite{Misra2017}.
All Lennard-Jones interactions were truncated and shifted at 
\SI{10}{\angstrom}. 
Electrostatic interactions in real space were cut off at \SI{10}{\angstrom} and 
long-ranged interactions were evaluated using particle--particle 
particle--mesh Ewald summation \cite{hockney1988} such that
the RMS error in the forces was a factor of $10^5$ smaller 
than the force between two unit charges separated by a 
distance of \SI{1.0}{\angstrom} \cite{Kolafa1992}.

The simulations were carried out in the canonical (NVT) ensemble, 
where the temperature was held at $\SI{300}{K}$.
Two separate Nos\'{e}--Hoover thermostats \cite{Shinoda2004,Tuckerman2006} 
were applied to each water slab, each being a Nos\'{e}--Hoover 
chain with 10 thermostats and a damping constant of
$\SI{0.1}{\pico\second}$.
The Langevin thermostat was used to control the dissipation of the 
Drudes particles in the solid, with a relaxation rate of $\gamma$, 
giving a random force $f_{\rm ran}\propto\sqrt{k_{\rm B}T m_{\rm D}\gamma}$.
Dynamics were propagated using the velocity Verlet algorithm with a 
time-step of $\SI{1}{\femto\second}$, unless specified otherwise. 
Each system was equilibrated for $\SI{100}{\pico\second}$ and the 
subsequent $\SI{10}{\nano\second}$ was used for analysis to give
the results presented in the main article.

\subsection{Computation of surface response functions and friction coefficients}

We characterise the charge density distributions of the
solid and the liquid by their surface response
functions defined as
\begin{equation}
\mathrm{Im}\,g(q,\omega) = 
\frac{\pi\omega}{q \mathcal{A} k_{\mathrm{B}} T}
\int_{-\infty}^\infty\!\mrm{d}t\,\mrm{e}^{i\omega t}
\sum_{\alpha,\beta}\!
\big\langle Q_{\alpha} Q_{\beta}\,
\mrm{e}^{-i\mbf{q}\cdot[\mbf{x}_{\alpha}(t)-\mbf{x}_{\beta}(0)]}
\mrm{e}^{-q |z_{\alpha}(t)-z_0|}\,
\mrm{e}^{-q |z_{\beta}(0)-z_0|}
\big\rangle.
\end{equation}
Here, $Q_{\alpha}$ is the charge on atom $\alpha$,
whose in-plane position at time $t$ is
$\mbf{x}_\alpha(t)$, $\mbf{q}$ is a wavevector parallel to the
graphene sheet, $z_\alpha(t)$ is the vertical coordinate and $z_0$ 
is a plane $1.6$\,\AA{} away from the edge of the water slab or solid surface. 
In practice, we computed at every time-step (\SI{1}{\femto\second})
the Fourier--Laplace surface components of the charge densities for
each water slab and each solid layer
\begin{equation}
\tilde{n}(q,t) = \sum_{\alpha}
Q_{\alpha}\mrm{e}^{i\mbf{q}\cdot\mbf{x}_{\alpha}(t)}
\mrm{e}^{-q|z_{\alpha}(t)-z_0|}.
\end{equation}
As we are interested in the long-wavelength limit
($q \rightarrow 0$), we focus on $\mbf{q}=\mbf{q}_0$, the lowest
wavevector in the $x$ direction accessible in our simulation box, the
magnitude of which is $q_0 =
2\pi/L_x \approx \SI{0.25}{\per\angstrom}$ where $L_x$ is the length
of the box in the $x$ direction.
The power spectra of the surface charge densities are given as
\begin{equation}
S(q,\omega)=\frac{1}{\mcl{A}}\int^{+\infty}_{-\infty}\!
\mrm{d}t \,\langle \tilde{n}(q,0)
\,\tilde{n}(-q,t)\rangle \,
\mrm{e}^{i\omega t}.
\end{equation}
Through the fluctuation-dissipation theorem, we can obtain the
imaginary part of the surface response function through
\begin{equation}
\mrm{Im}\,g(q,\omega) = 
\frac{2\pi}{q}\frac{\omega}{2k_{\rm B}T}S(q,\omega).
\end{equation}
To ensure the spectrum is independent of noise, a Savitzky--Golay 
filter \cite{Savitzky1964} was applied.
The resulting spectra without further fitting are presented in the
main article.

The liquid--solid friction coefficient is computed via the Green--Kubo
relation in analogous manner to the previous section. For a system with
$N=3$, we show the Green--Kubo integral $\lambda_{\rm GK}$ in Fig.~2D and
the corresponding total friction coefficient $\lambda$ in Fig.~2E.
Since in our simulations, changing $m_{\rm D}$ does not affect static
equilibrium properties such as surface roughness, we can extract
the surface-roughness classical friction contribution 
$\lambda_{\rm cl}\approx2.2\times10^4\,\mrm{N\,s\,m}^{-3}$
on the water slab A from the 
simulations with $\omega_{\rm p}>20\,\mrm{THz}$. This is only
marginally higher than $\lambda_{\rm cl}\approx1.9\times10^4\,\mrm{N\,s\,m}^{-3}$ 
estimated in Ref.~\onlinecite{Bui2023} for a single water slab 
on a single graphene sheet, with the difference coming from 
the presence of two extra graphene layers and water slab B underneath.

\subsection{Non-equilibrium simulation details}

\begin{figure*}[t]
 \centering \includegraphics[width=0.75\linewidth]{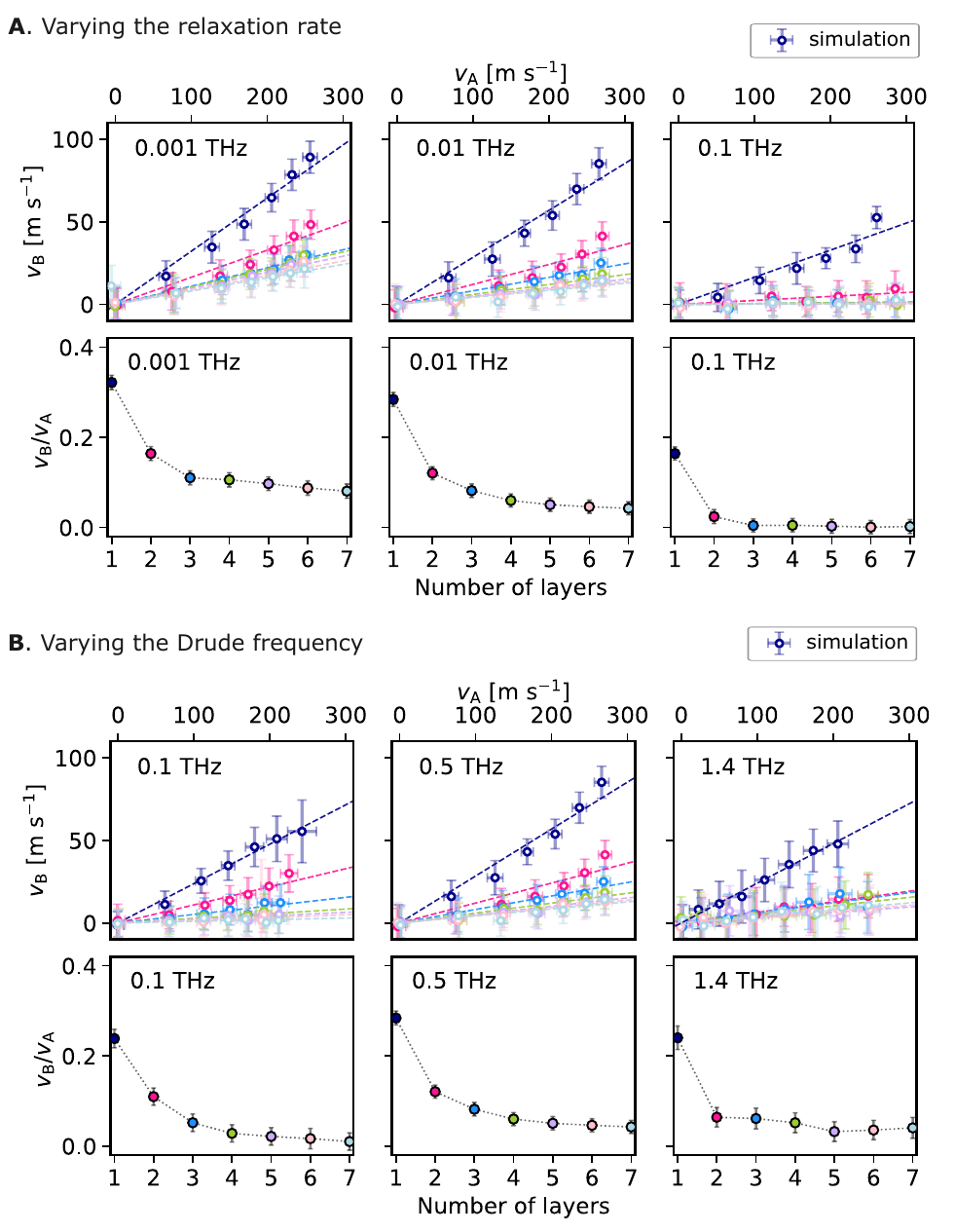} 
 \caption{\textbf{Flow tunneling efficiency from non-equilibrium
 simulations with active walls.}
 \tb{(A)} Results for different relaxation rates of the Drude particles'
 dissipation $\gamma$ (as indicated in the left corner of each plot)
 for a Drude frequency of $\omega_{\rm p}=0.5\,\mrm{THz}$.
   The different colors on the top line correspond to the different number of layers as represented on the bottom line. The values in the bottom panel correspond to the gradient of the fitted (dashed) curves in the top panel.
 \tb{(B)}  Results for different Drude frequencies $\omega_{\rm p}$
  (as indicated in the left corner of each plot) for a 
  relaxation rate of $\gamma=0.01\,\mrm{THz}$.
  The different colors in the top panel correspond to the different number of layers as represented on the bottom panel. The values in the bottom panel correspond to the gradient of the fitted (dashed) curves in the top panel.
 } 
 \label{SI_fig3}
\end{figure*}

Non-equilibrium molecular dynamics (NEMD) simulations 
are carried out to study pressure-driven flow. To drive 
the flow of the top water film, we applied an external force 
on the O atom of each water molecules in the $x$ direction
while keeping the walls immobile, which generated a
Poiseuille flow. The thermostats are applied only after excluding the 
center-of-mass contribution with a damping time of $100\,\mrm{fs}$.

To compute the flow-tunneling efficiency for a set-up
with a particular number of graphene layers $N$, relaxation
rate $\gamma$ and Drude frequency $\omega_{\rm p}$,
at least ten different simulations were run with different
magnitudes of the external force. The corresponding friction
force ranges up to $F/\mathcal{A}\sim25\times10^5\,\mrm{N\,m^{-2}}$ 
for simulations with $m_{\mrm{D}}=5\,\mrm{amu}$ and 
to $F/\mathcal{A}\sim500\times10^5\,\mrm{N\,m^{-2}}$ 
for simulations with $m_{\mrm{D}}=500000\,\mrm{amu}$.
After at least $1\,\mrm{ns}$ of equilibration,
the velocity of both water slabs are computed by
averaging over $\sim10\,\mrm{ns}$ of production run.
A linear regression line is fitted to the induced velocity 
$v_{\mrm{B}}$ as function to the driven velocity $v_{\rm A}$
using least-squares fitting. The ratio $v_{\mrm{B}}/v_{\mrm{A}}$ 
is given as the slope of this regression line, with
the error given as the standard error of the slope.
The results for a subset of simulations are shown in Fig.~3.

\section{Theoretical setup}

\begin{figure*}[t]
 \centering \includegraphics{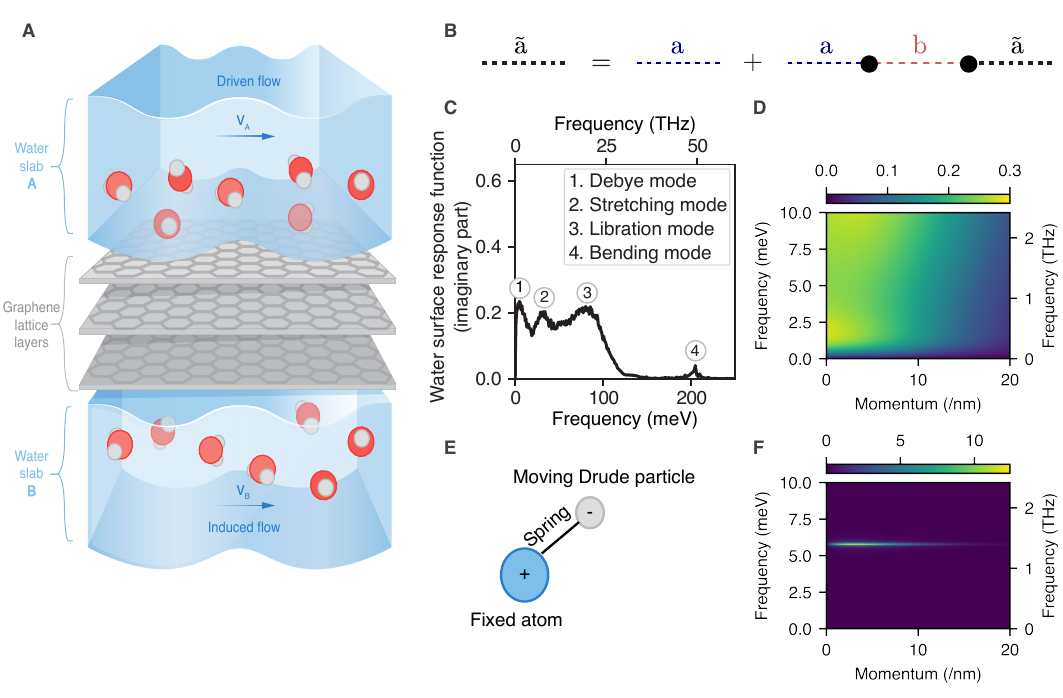} 
 \caption{\textbf{Model.}
 \tb{(A)} Schematic illustrating the system with two liquid films separated by a multi-layer solid. The top liquid is driven out-of-equilibrium and moves at velocity $\v_{\rm A}$, transferring momentum to the electrons in the solid via Coulomb drag. After crossing the $N$ layers of solid, this induced excitation further couples to and drags the liquid at the bottom which reaches a velocity $\v_{\rm B}$.  
  \tb{(B)} Dyson equation to compute the Green's function of the layer a renormalized by the layer b. Here, the propagators $\tilde{a}$ represents the renormalized Green's function of  layer a, while the propagators a (resp. b) represent the bare Green's functions of layer a (resp. b).
  \tb{(C)}Spectrum of water ($\im{g_{\rm w}}$) as a function of frequency at momentum $q=6.7$ nm\1, obtained from simulations. The different peaks correspond to different excitation modes (Debye, stretching, libration, bending).  
  \tb{(D)} Water response function $\im{g_{\rm w}}$ as a function of both frequency and momentum using the fit Eq. \eqref{gw}.  
   \tb{(E)} Schematic representation of the Drude oscillators used to model the electronic modes. An effective electron is attached to an effective nucleus by a spring, generating a plasmon mode. 
   \tb{(F)} Drude oscillators response function $\im{g_{\rm Dr}}$ as a function of both frequency and momentum using the formula of Eq. \eqref{gdrude}.  Here, we used $d=1.675$ \AA, a Drude frequency $\omega_{\rm p}=1.4$ THz and a relaxation rate $\gamma=0.1$ THz. 
 } 
 \label{SI_fig4}
\end{figure*}

\subsection{Model}

We consider two slabs of water, A and B, filling two half-spaces separated by $N$ two-dimensional solid layers, as pictured in Fig. \ref{SI_fig4}A. These layers are assumed translationally invariant in the $(x, y)$ plane. We denote $n_i$ the charge density in the $i$-th layer, with $i=0=$~A being the first water slab and $i=N+1=$~B the second water slab. Each layer is neutral, thus $\langle n_i\rangle=0$. However, there can be fluctuations of charges. 

 Each layer is described by a Hamiltonian $\Ha_i$ which takes into account the internal structure. In addition, the charges interact electrically both inside a layer and between layers. These interactions are of the form:
\beq \Ha_{ij}(t)=\int\dd\x\dd\x'\, n_i(\x,t)V(\x-\x')n_j(\x',t)\eeq
where $V(\x)=e^2/(4\pi\epsilon_0 r)$ is the Coulomb potential. In the following we treat the inter-layer electric interactions as a perturbation. We consider interactions only between nearest-neighbor layers.

\subsection{Linear response theory}

In the perturbative framework, we will be interested by the response of a layer to the electric potential of another layer. Let us consider fluctuation $\delta n_i(\x,t)$ of the charge density in the layer $i$. This fluctuation generates an electric potential 
\beq \delta \phi_i(\x,t)=\int\dd\x'\, V(\x-\x')\delta n_i(\x',t)\eeq
The charge density induced in layer $j$ in response to this potential is, within linear response theory:
\beq \delta n_j(\x,t)=\int_{-\infty}^t\dd t'\, \langle [n_j(\x,t),\Ha_{ij}(t')] \rangle=\int_{-\infty}^t\dd t' \int\dd\x'\dd\x''\,  \langle[n_j(\x,t), n_j(\x',t')] \rangle V(\x'-\x'')\delta n_i(\x'',t').\eeq
In order to simplify these convolutions, we define a transformed charge density: 
\beq \bar{n}_i(\q,t)=\sqrt{\frac{e^2}{2\epsilon_0 q}}\int\dd \x\, n_i(\x,t)e^{-i\q\cdot\x-q|\Delta z|}\eeq
where $\q$ is the in-pane momentum and $\Delta z$ the distance to the interface. Here, the integral over $z$ disappears for a 2d layer. 
With this definition,  the interaction Hamiltonian reduces to 
\beq \Ha_{ij}(t)=\int\frac{\dd\q}{(2\pi)^2}\, \bar{n}_i(\q,t)^*\bar{n}_j(\q,t)\eeq
and the linear response of the layer $j$ to a charge fluctuation in the layer $i$ becomes
\beq \delta \bar{n}_j(\q,\omega)=- g_j^{\rm R,0}(\q,\omega)\delta \bar{n}_i(\q,\omega)\eeq
where 
\beq g_j^{\rm R,0}(\tb{r},t,\tb{r}',t')= i\theta(t-t')\langle[\bar{n}_j(\tb{r},t),\bar{n}_j(\tb{r}',t')]\rangle\eeq
is the bare surface response function of the layer $j$, corresponding to $(-)$ the usual retarded response function for the effective density $\bar{n}_j$. 

Taking into account the self-screening inside the layer $j$ at the RPA level, the surface response function becomes 
\beq g_j^{\rm R}(\q,\omega)=\frac{g_j^{\rm R,0}(\q,\omega)}{1+g_j^{\rm R,0}(\q,\omega)}.\eeq

To formalise these definitions and develop a rigorous perturbation theory, we introduce the Keldysh formalism. 

\subsection{Keldysh formalism}

In the following, we use the Keldysh formalism, which is a non-equilibrium perturbation theory. We start by stating some elements of this formalism \cite{Rammer2007}. 

We consider particles (bosons or fermions) with creation and annihilation operators $\psi^\dagger(\tb{r},t)$ and $\psi(\tb{r},t)$, respectively. We describe the system's dynamics in terms of three types of real-time Green's functions: the Retarded, Advanced and Keldysh Green's functions, defined, for both bosons and fermions, according to 
\beqa\label{definition_Green}
\left\{ \begin{array}{l}
G^{\rm R}(\tb{r},t,\tb{r}',t')= -i\theta(t-t')\langle[\psi(\tb{r},t),\psi^\dagger(\tb{r}',t')]_s\rangle,\\
G^{\rm A}(\tb{r},t,\tb{r}',t')= i\theta(t'-t)\langle[\psi(\tb{r},t),\psi^\dagger(\tb{r}',t')]_s\rangle,\\
G^{\rm K}(\tb{r},t,\tb{r}',t')=-i\langle[\psi(\tb{r},t),\psi^\dagger(\tb{r}',t')]_{-s}\rangle,
\end{array}\right.
\eeqa 
where $\psi^{\dagger}$ and $\psi$ are the particles' creation and annihilation operators, and $[A,B]_\pm=AB\pm BA$, $s$ being $+$ for fermions and $-$ for bosons. In the following, we restrict the discussion to bosons. 

The Retarded and Advanced Green's functions contain information on the system's elementary excitations. For free bosons (such as plasmons) with dispersion $\omega_{\q}$ in a translationally-invariant system at equilibrium, the Fourier-transformed Green's functions are 
\beq
G^{\rm R,A} = \frac{\omega_\q}{(\omega\pm i0^+)^2-\omega_\q^2}. 
\eeq
the Keldysh Green's function contains information on the quasiparticle distribution. At equilibrium, it satisfies the fluctuation-dissipation theorem: 
\beq\label{DFD}
  G^{\rm K}(\tb{q},\omega)=2 i\,\coth\left(\frac{\hbar\omega}{2T}\right)\im{G^{\rm R}(\tb{q},\omega)}.
  \eeq
   Thus, at equilibrium, 
  \beq
  G^{\rm K} = (2n_{\rm B}(\omega) +1) \times   i \pi \left[\delta(\omega- \omega_\q)-\delta(\omega+ \omega_\q)\right], 
  \eeq
 where we recover the Bose-Einstein distribution $n_{\rm B}(\omega) = 1/(e^{\hbar\omega/T} - 1)$. 
 The Keldysh Green's functions are therefore the analogues of the occupation distribution functions in the approximate Boltzmann formalism. They will be key in determining the non-equilibrium state of the system.
 
 Here, the system is driven out-of-equilibrium by the flow at velocity $\v_{\rm A}$ imposed in water slab A. Therefore, the water Green's functions have the form
 \beq G(\tb{r},t,\tb{r}',t')=G_{\rm static}(\tb{r}-\v_{\rm A}t,t,\tb{r}'-\v_{\rm A}t',t')\eeq
 where $G_{\rm static}$ is the Green's of the static slab of water. Going to Fourier space, this drift becomes a frequency shift:
  \beq G(\q,\omega)=G_{\rm static}(\q,\omega-\q\cdot\v_{\rm A}).\eeq
 Let us outline that these Green's functions no longer fulfill the fluctuation-dissipation theorem: the system is out-of-equilibrium, justifying the use of the Keldysh formalism. Indeed, we now find:
\beq
  G^{\rm K}(\tb{q},\omega)=2 i \, \coth\left(\frac{\hbar\omega-\hbar\q\cdot \v_{\rm A}}{2T}\right)\im{G^{\rm R}(\tb{q},\omega)}.
  \eeq
  
Finally, because of the flow tunneling, the induced flow in slab B reaches a steady state velocity $\v_{\rm B}$. Similarly, we model this effect by a frequency shift in the Green's function of slab $B$. 

\subsection{Dyson equation}

Our task is now to compute the non-equilibrium Green's functions of the layer $i+1$ in the presence of the perturbation applied by the layer $i$. This computation is the bosonic analogue to the computation of the electron drag in \cite{Coquinot2023}. In the Keldysh formalism, we consider the matrix Green's function 
\beqa\label{definition_Green0}
\tb{G}=\left( \begin{array}{cc}
G^{\rm R} & G^{\rm K}\\
0 & G^{\rm A}
\end{array}\right).
\eeqa 
The perturbation series can be exactly resummed by the non-equilibrium Dyson equation which is represented diagrammatically in Fig. \ref{SI_fig4}b:
\beq
\tilde{\tb{G}}_{i+1} = \tb{G}_{i+1} + \tb{G}_{i+1} \otimes \tb{G}_{i} 
\label{dyson}
\eeq
where $\tb{G}_{i}$ is the non-interacting Green's function and  $\tilde{\tb{G}}_{i}$ the renormalized Green's function. Here, $\otimes$ represents convolution in space and time through the Coulomb potential, as well as matrix multiplication:
\beq (\tb{A} \otimes  \tb{B} )_{ab}(\x,t,\x',t') = \sum_c\int\dd \x_{\rm A}\x_{\rm B}\dd t'' \, A_{ac}(\x,t,\x_{\rm A},t'')V(\x_{\rm A}-\x_{\rm B})B_{cb}(\x_{\rm B},t'',\x',t') \eeq
 We assume that the system is translationally invariant parallel to the interface, and that it has reached a steady state: we may then Fourier-transform Eq.~\eqref{dyson}. If the layer is not 2d, we also need to integrate in the $z$-direction. Finally, to simplify the computations, we include the Coulombic pre-factor into the Green's functions, leading to introduce the surface response function:
 \beq g(\q,\omega)=-\frac{e^2}{2\epsilon_0 q}\int\dd z\dd z'\, G(\q,z,z',\omega)e^{-q(|z|+|z'|)}\eeq
 where the integration over $z$ disappears for 2d layers and the origin of $z$ corresponds to the interface. 
 
  With the convolutions becoming products in Fourier space, and using that $G^{\rm A}(\q,\omega) = G^{\rm R} (\q,\omega)^*$,
\beqa\label{G2R}
\tilde{g}^{\rm R}_{i+1}(\q,\omega) &=& \dfrac{g_{i+1}^{\rm R} - |g_{i+1}^{\rm R}|^2 \left(g^{\rm R}_{i}\right)^*}{\left|1- g_{i+1}^{\rm R} g^{\rm R}_{i}\right|^2},\\
\tilde{g}^{\rm K}_{i+1}(\q,\omega) &=& \dfrac{g_{i+1}^{\rm K} + |g_{i+1}^{\rm R}|^2 g^{\rm K}_{i}}{\left|1- g_{i+1}^{\rm R} g^{\rm R}_{i}\right|^2}. 
\eeqa

For the multi-layer solid with nearest-neighbor interactions, we can compute the surface response function of layer $i+1$ renormalized by the $i$ previous layers of solid, plus the driven flow, through an iterative procedure. Indeed, we the renormalise the response function of layer $i+1$ by the response function of  layer $i$, itself renormalized by the Green's function of  layer $i-1$, until we reach the driven flow ($i=0$). As a consequence, we can compute the response function of the $N$-th layer of solid driven out-of-equilibrium by the driven flow through the $N-1$ previous layers of solid. 

\subsection{Out-of-equilibrium friction forces}

We can compute the force between the layer $i$ and the layer $i+1$ when the system is out-of-equilibrium. This force is formally
\beq
\tb{F}_{i\to i+1} = \int \dd \x_i \dd \x_{i+1} \langle n_i(\x_i, t) V(\x_i - \x_{i+1}) n_{i+1}(\x_{i+1}, t)\rangle, 
\label{basic_force}
\eeq
Thus, we need to compute a density-density correlation function, which can be done in Keldysh perturbation theory. Indeed, according to \cite{Kavokine2022}, this force writes:
\beq \frac{\tb{F}_{i\rightarrow i+1}}{\A}=\frac{1}{2}\int\frac{\dd\q\dd\omega}{(2\pi)^3} (\hbar\q)\, \frac{g_i^{\rm R}(\q,\omega)g_{i+1}^{\rm K}(\q,\omega)+g_{i}^{\rm K}(\q,\omega)g_{i+1}^{\rm A}(\q,\omega)}{|1-g_{i}^{\rm R}(\q,\omega)g_{i+1}^{\rm R}(\q,\omega)|^2}\eeq
Using that $\tb{F}_{i\rightarrow i+1}=-\tb{F}_{i+1\rightarrow i}$, we deduce
\beq \frac{\tb{F}_{i\rightarrow i+1}}{\A}=-\frac{1}{2i}\int\frac{\dd\q\dd\omega}{(2\pi)^3} (\hbar\q)\, \frac{\im{g_{i}^{\rm R}(\q,\omega)}g^{\rm K}_{i+1}(\q,\omega)-g^{\rm K}_i(\q,\omega)\im{g_{i+1}^{\rm R}(\q,\omega)}}{|1-g_i^{\rm R}(\q,\omega)g_{i+1}^{\rm R}(\q,\omega)|^2} \label{generic_friction}
\eeq
This formula gives the force between two systems interacting with a density-density interaction in general, even when they are out-of-equilibrium. As expected, it has the form of a Landauer formula:
\beqa 
\frac{\tb{F}_{i\rightarrow i+1}}{\A}&=&\int\frac{\dd\q}{(2\pi)^2} (\hbar\q)  (\Gamma_{i\rightarrow i+1}(\q) - \Gamma_{i+1\rightarrow i}(\q))
\label{generic_friction}
\eeqa
with
\beqa 
 \Gamma_{i\rightarrow i+1}(\q)=\int\frac{\dd\omega}{4i\pi} \frac{\im{g_{\rm i+1}^{\rm R}(\q,\omega)}g^{\rm K}_{i}(\q,\omega)}{|1-g_{\rm i}^{\rm R}(\q,\omega)g_{\rm i+1}^{\rm R}(\q,\omega)|^2}.
\eeqa
Physically, $g^{\rm K}_{i}(\q,\omega)$ counts the number of quasi-particles in the layer i at momentum $\hbar\q$ and energy $\hbar\omega$, while $\im{g_{\rm i+1}^{\rm R}(\q,\omega)}$ is the "density of states" in the layer i+1 at the corresponding momentum and energy to which a quasi-particle may tunnel.


Finally, to compute the friction force of the layer $i$ on the layer $i+1$, we can apply this formula where the Green's function of both layers are renormalized. Because we are computing the force originating from the interaction $\Ha_{i, i+1}$, the Green's function are to be renormalized by every interaction except this one -- otherwise the interaction would be counted twice. Therefore, the layer $i$'s Green's function is to be renromalised by the layer $i-1$ then $i-2$ until $0$ (driven flow) while the layer $(i+1)$'s layer is to be renormalized by the layer $i+2$ then $i+3$ until $N+1$ (induced flow).

\section{Models of water and solid}

\subsection{Water surface response function and hydrons}

The surface response function of water has been studied numerically with Molecular Dynamics (MD) simulations \cite{Bonthuis2012,Kavokine2022,Coquinot2023b}. It contains several "hydron" modes (Fig. \ref{SI_fig4}C); at the thermal frequencies that are relevant for quantum friction, those are mainly the Debye mode and the libration mode. Following \cite{Kavokine2022}, we describe the water response function by the following formula: 
\beq g_{\rm w}^{\rm R}(\q,\omega)=\frac{1}{2}\left[\frac{\omega_{\rm D,1}}{\omega_{\rm D,1}-i\omega}e^{-q/q_0}+\frac{\omega_{\rm D,2}}{\omega_{\rm D,2}-i\omega}\left(2-e^{-q/q_0}\right)\right]e^{a+a'(1+(q/b)^\alpha)^{1/\alpha}}
\label{gw}
\eeq
where the Debye frequencies are $\omega_{\rm D,1}=0.36$ THz (\ie $\hbar\omega_{\rm D,1}=1.5$ meV) and $\omega_{\rm D,2}=4.8$ THz (\ie$\hbar\omega_{\rm D,2}=20$ meV), $q_0=3.12$ \AA\1, $a=5.16$, $a'=-5.19$, $b=1.95$ \AA\1 and $\alpha=2$. In particular, the dispersion is weak. This formula is plotted in Fig. \ref{SI_fig4}D. 


Within our theoretical framework, we further need to specify the position of a boundary plane separating the water from the solid, which is essentially set by the static water-solid interaction parameters. Withe the Lennard-Jones parameters of carbon atoms, refs.~\cite{Kavokine2022, Bui2023} found that this plane is located at a distance $\delta = 1.3$~\AA from the atomic centers. In our passive wall simulation, agreement with the theory was obtained by setting $\delta = 1$~\AA. 


\subsection{Drude model of solid}

We consider a lattice of Drude oscillators at $z=0$ whose electron has a mass $m_{\rm Dr}$ and a charge $-Q_{\rm Dr}$. Each electron is attached to a fixed ion of charge $Q_{\rm Dr}$ through a spring of stiffness $K_{\rm Dr}$, as sketched in Fig. \ref{SI_fig4}e. We denote $\x_k^0$ the position of the $k$-th ion and $\x_k$ the distance between the $k$-th electron and the $k$-th ion, so that the harmonic force between them writes 
\beq \tb{F}_{\rm ho}^k=-K_{\rm Dr}\x_k.\eeq
The oscillator is damped by a drag force
\beq \tb{F}_{\rm th}^k=-2m_{\rm Dr}\gamma\partial_t \x_k\eeq
which corresponds in the simulation to the action of a thermostat through and physically the the effect of phonons and impurities. 
The only interaction between oscillators are electric. Each oscillator is affected by the local electric potential $\phi$. To compute the response function we apply an external electric potential $\phi_{\rm ext}$. The local electric field is the sum of an external field and the field induced by the oscillators.

The momentum balance of the $k$-th oscillator writes 
\beq m_{\rm Dr}\partial_t^2 \x_k=-K_{\rm Dr}\x_k-2m_{\rm Dr}\gamma\partial_t \x_k+Q_{\rm Dr}\nabla\phi(x_k,t). \eeq

We solve this model in the continuum limit. We denote $\tb{u}(\x_k^0,t)=\x_k(t)$ the displacement field and then obtain 
\beq m_{\rm Dr}\partial_t^2 \tb{u}=-K_{\rm Dr}\tb{u}-2m_{\rm Dr}\gamma\partial_t \tb{u}+Q_{\rm Dr}\nabla\phi(x,t). \eeq
Going to Fourier space we obtain
\beq -m_{\rm Dr}\omega^2 \tb{u}=- K_{\rm Dr} \tb{u}+2im_{\rm Dr}\gamma\omega \tb{u} +iQ_{\rm Dr} \q\phi(\q,\omega) \eeq
that is 
\beq \tb{u}(\q,\omega)=\frac{iQ_{\rm Dr} \q}{ K_{\rm Dr}  -m_{\rm Dr}\omega^2-2im_{\rm Dr}\gamma\omega  }\phi(\q,\omega). \eeq

This displacement field $\tb{u}(\x,t)$ generates an electric potential. In the continuum limit, we denote $\rho$ the 2d density of Drude oscillators. At that point, we should distinguish between the longitudinal and transverse modes. For simplicity, we restrict the discussion to the former, the latter being similar. Thus, the induced potential is 
\beq \phi_{\rm ind}(\tb{r},z,t)=\int\rho\dd^2\tb{r}'\, Q_{\rm Dr}[V(\tb{r}-\tb{r}',z)-V(\tb{r}-\tb{r}'-\tb{u}(\tb{r}',t),z)]\approx \int\rho\dd^2\tb{r}'\, Q_{\rm Dr}\nabla V(\tb{r}-\tb{r}',z)\cdot \tb{u}(\tb{r}',t)\eeq
where $V$ is the Coulomb potential. Going to Fourier space, we obtain, 
\beq \phi_{\rm ind}(\tb{q},z,\omega)=\frac{\rho Q_{\rm Dr}}{2\epsilon_0q}e^{-q|z|}i\tb{q} \cdot \tb{u}(\tb{q},\omega)=-\frac{\rho Q_{\rm Dr}^2}{2\epsilon_0 m_{\rm Dr}}\frac{ qe^{-q|z|}}{ \frac{ K_{\rm Dr} }{m_{\rm Dr}} -\omega^2-2i\gamma\omega  }\phi(q,\omega)\eeq
where $n_{\rm Dr}(\x,t)=\rho Q_{\rm Dr}\nabla\cdot\tb{u}(\x,t)$ is the charge density in the material. 
  Denoting $\omega_{\rm p}= \sqrt{K_{\rm Dr} /m_{\rm Dr}}$ the Drude frequency and $\ell_{\rm p}=\rho Q_{\rm Dr}^2/(2\epsilon_0 K_{\rm Dr})$ a length, we have 
\beq g_{\rm Dr}^{\rm R}(\q,\omega)=\frac{ \omega_{\rm p}^2\ell_{\rm p} q}{ \omega_{\rm p}^2 -\omega^2-2i\gamma\omega  }, \qquad  \phi_{\rm ind}(\tb{q},z,\omega)=-g_{\rm Dr}^{\rm R}(\q,\omega)e^{-q|z|}\phi(q,\omega).\eeq
At this stage, we have not taken into account the Coulomb interaction between the Drude oscillators. Including them would lead to a dispersion of the Drude frequency. However, the simulations show that this dispersion is negligible, and we therefore use the non-interacting expression in the following. 


Since we always consider interactions between layers separated by a distance $d_0$, it will be convenient to redefine the surface response function according to 
\beq g_{\rm Dr}^{\rm R}(\q,\omega)=\frac{\omega_{\rm p}^2\ell_{\rm p} qe^{-2qd_0}}{\omega_{\rm p}^2 -\omega^2-2i\gamma\omega }, 
\label{gdrude}
\eeq
and then carry out the computations as if there was no separation between the layers. We use we use $d_0=1.675$ \AA\, (half the distance between two graphitic layers) when computing the response to another solid layer, and $d_0 = \delta$ when computing the response to a water slab. Eq.~\eqref{gdrude} is plotted in Fig. \ref{SI_fig4}F.


\subsection{Effect of phonons}

 The effect of phonons on flow tunneling was previously studied by Andreev and Meierovitch \cite{Andreev1971} and shown to be negligible for all practical purposes. In their model, the Authors consider the flow induced throughout a dissipation-less solid induced by a Poiseuille flow in a channel of height $a$. Acording to their results, the induced velocity writes:
 \beq \frac{v_B}{v_A}=\zeta\left(\frac{3}{2}\right)
 \frac{T}{4\eta}
 \sqrt{\frac{\rho_\ell^5c_\ell^3}{\pi^5\rho_s^2y^3ac_s^T}}
 \Phi\left(\frac{c_\ell}{c_s^T},\frac{c_s^L}{c_s^T}\right)\label{Andreev}
 \eeq
 where $\zeta$ is the Riemann-function and $\Phi$ is a dimensionless pre-factor defined in \cite{Andreev1971} and typically ranging between $10^{-2}$ and 1. Here, $\rho_\ell$ and $\rho_s$ are the densities of the liquid and the solid respectively, $c_\ell$ is the velocity of sound in the liquid and $c_s^L$ and $c_s^T$ the velocity of longitudinal/transverse sound in the solid. 
Finally, $y=4\eta/3+\hat\eta+\kappa(c_p-c_v)/c_pc_v$ is a parameter of the liquid with $\kappa$ the thermal conductivity $c_p$ and $c_v$ the specific heats per unit mass, and $\eta$ and $\hat \eta$  the first and second viscosities. 

Notably, the efficiency of the phonon-mediated flow tunneling scales with $1/\sqrt{a}$. Thus, the flow tunneling is larger when the driving flow is confined. Physically, this comes from the stronger shear in the confined Poiseuille flow which produces more phonons. However, the model of \cite{Andreev1971} neglects slippage and then the coefficient $a$ in Eq. \eqref{Andreev} should be replaced by the slip length at strongest confinements. Finally, applying Eq. \eqref{Andreev} to nano-confined liquids separated by a graphite wall, one finds a tunneling efficiency $v_{\rm B}/v_{\rm A} \sim 10^{-5}$, which is negligible.

\section{Flow tunneling}

\subsection{Simplified expression of fluctuation-induced friction}

Before considering the flow tunneling geometry, we first provide a simplified expression of the solid-liquid quantum friction force, starting from Eq.~\eqref{generic_friction}. We assume the liquid, labeled w, to flow at velocity $\v$ while the solid, labelled e, is at equilibrium. Then, 
\beq g_{\rm w}^{\rm K}(\q,\omega)=\coth\left(\frac{\hbar\omega-\hbar \q\v}{2T}\right) \im{g_{\rm w}^{\rm R}(q,\omega)}, \qquad  g_{\rm e}^{\rm K}(q,\omega)=\coth\left(\frac{\hbar\omega}{2T}\right) \im{g_{\rm e}^{\rm R}(q,\omega)}. \eeq
Thus, to first order in $\v$, and after carrying out the angular integration, Eq. \eqref{generic_friction} becomes 
\beq \frac{\tb{F}_{\rm e\rightarrow w}}{\A}=-\frac{\hbar^2}{16\pi^2T}\tb{v}\int_0^\infty\dd q\, q^3\int_{-\infty}^\infty\dd \omega  \frac{1}{\sh^2\left(\frac{\hbar\omega}{2T}\right)}  \frac{\im{g^{\rm R}_{\rm w}(q,\omega)}\im{g_{\rm e}^{\rm R}(q,\omega)}}{|1-g_{\rm w}^{\rm R}(q,\omega)g_{\rm e}^{\rm R}(q,\omega)|^2},
\label{QF_formula}
\eeq
which is the result of \cite{Kavokine2022}. 

For the Drude solid in the limit of small damping ($\gamma\to0$), the surface response function  becomes 
\beq \im{g^{\rm R}_{\rm Dr}(q,\omega)} = \frac{\pi}{2}\omega\,\delta(\omega\pm\omega_p)\times q\ell_pe^{-2qd_0}. \eeq
Here, $d_0 = \delta = 1.3$ \AA \, is the distance between the atomic centers and the imaginary plane separating water from the solid (see Sec. IV A). Further neglecting the denominator (that corresponds to higher order corrections) in Eq.~\eqref{QF_formula} and approximating $\hbar\omega_p\ll T$ in the integrand, the quantum friction coefficient becomes
\beq \lambda_{\rm qf}=\frac{T\ell_p}{4\pi}\int_0^\infty\dd q\, q^4e^{-2qd_0}\times\frac{1}{\omega_p}\,  \im{g^{\rm R}_{\rm w}(q,\omega_p)}
\eeq
We may further neglect the momentum dependence of the water surface response function and approximate it as a single Debye peak, that is
\beq \im{g_{\rm w}^{\rm R}(\q,\omega)}\approx \alpha_{\rm w}\frac{\omega_0\omega}{\omega_0^2+\omega^2},\eeq
where $ \alpha_{\rm w}$ is an effective amplitude of the Debye peak at around $q\sim 1/d_0$. Then, the quantum friction coefficient writes
\beq \lambda_{\rm qf}\approx\frac{3}{16\pi}\frac{T\ell_p\alpha_{\rm w}}{d_0^5}\frac{\omega_0}{\omega_0^2+\omega_p^2}\approx \frac{10^5\, \tn{N.s/m}^3}{1+(\omega_p/\omega_0)^2}.
\label{QF_simple_estimate}
\eeq
Thus, when the Drude plasmon frequency is larger than the Debye frequency, the quantum friction coefficient scales with $ \lambda_{\rm qf}\propto 1/\omega_p^2$, while the coefficient goes to a constant when the Drude plasmon frequency becomes small compared to the Debye frequency.  This simple formula provides a good fit to the full result of Eq.~\eqref{QF_formula}, with parameters $\alpha_{\rm w}\approx 2\e{-2}$ and the effective Debye frequency $\omega_0\approx 0.3$ THz (see Fig. \ref{SI_fig6}A).

Here, we discussed the case of a single interface between a slab of water and a Drude layer. For the scenario of flow tunneling with $N$ layers of solid before another slab of water at equilibrium, the quantum friction coefficient can also be computed and corresponds to the blue curves in \ref{SI_fig6}A. When $N=1$, the presence of water on the other side of the solid layer increases the quantum friction, as it brings about additional charge fluctuations. For larger values of $N$, the effect of the second water slab becomes negligible and the quantum friction is almost identical to the single layer case. 

\subsection{Flow tunneling through a passive solid}

\begin{figure*}
 \centering \includegraphics[width=0.95\textwidth]{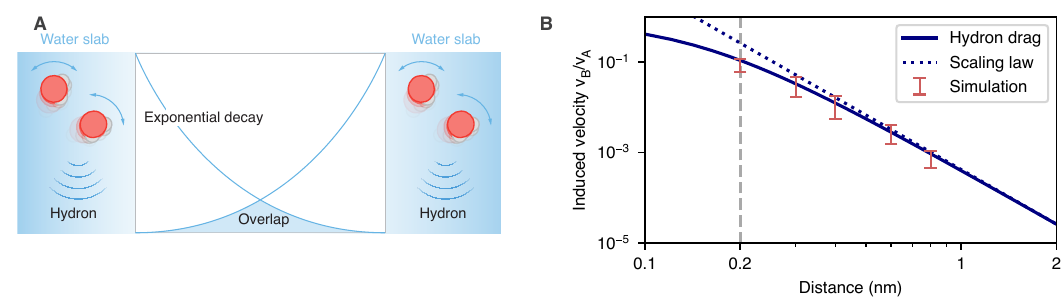} 
  \caption{\textbf{Flow tunneling through an inert solid}.
 \tb{(A)} Schematic of the tunneling of elementary excitations (hydrons) between two water slabs. Tunneling occurs when the associated Coulomb potentials overlap. 
 \tb{(B)} Ratio of the tunneling-induced velocity $v_B$ and imposed velocity $v_A$ versus thickness of the inert solid $d$, assumed to have a classical (roughness-induced) friction coefficient $\lambda_{\rm cl}\approx 2.1\e{4}~\rm N\cdot s \cdot m^{-3}$. The vertical dashed line indicates the value of $d$ below which the separation between the water slabs becomes ill-defined. Error bars represent the mean-squared error over simulations. 
 }  \label{SI_fig5}
\end{figure*}

We start with the case of an inert solid. According to Eq. \eqref{generic_friction} the force applied by water slab A on water slab B is 
\beqa 
\frac{\tb{F}_{\rm hh}}{\A}&=&\int\frac{\dd\q}{(2\pi)^2} (\hbar\q)  (\Gamma_{\rm A\rightarrow B}(\q) - \Gamma_{\rm B\rightarrow A}(\q))
\label{Fhh}
\eeqa
with
\beqa 
 \Gamma_{a\rightarrow b}(\q)=\int\frac{\dd\omega}{4i\pi} \frac{g^{\rm K}_{\rm w}(\q,\omega-\q\cdot\v_a)\im{g_{\rm w}^{\rm R}(\q,\omega-\q\cdot\v_b)}e^{-2qd}}{|1-g_{\rm w}^{\rm R}(\q,\omega-\q\cdot\v_a)g_{\rm w}^{\rm R}(\q,\omega-\q\cdot\v_b)e^{-2qd}|^2}
\eeqa
where $g_{\rm w}$ is the equilibrium water surface response function and $d$ is the thickness of the solid. Shifting the frequencies and using the fluctuation-dissipation theorem (Eq. \eqref{DFD}) leads to 
\beq  \Gamma_{a\rightarrow b}(\q)=\int\frac{\dd\omega}{2\pi} (2n_{\rm B}(\omega-\q\cdot\v_a)+1)\frac{\im{g_{\rm w}^{\rm R}(\q,\omega-\q\cdot\v_a)}\im{g_{\rm w}^{\rm R}(\q,\omega-\q\cdot\v_b)}e^{-2qd}}{|1-g_{\rm w}^{\rm R}(\q,\omega-\q\cdot\v_a)g_{\rm w}^{\rm R}(\q,\omega-\q\cdot\v_b)e^{-2qd}|^2}\eeq
Here, $f^{a}_{\q}(E)=n_{\rm B}(E/\hbar-\q\cdot\v_a)$ is the average number of hydrons of wavevector $\q$ and energy $E$.
Using that $\re{g_{\rm w}^{\rm R}(\q,\omega)}$ is symmetric in $\omega$ and $\im{g_{\rm w}^{\rm R}(\q,\omega)}$ is antisymmetric in $\omega$, Eq. \eqref{Fhh} becomes, to first order in velocities:
\beqa 
\frac{\tb{F}_{\rm hh}}{\A}&=&\frac{1}{2 \pi \hbar}  \int \dd E\int\frac{\dd\q}{(2\pi)^2} (\hbar\q)  (f^{T}_{\q}(E)-f^{B}_{\q}(E)) \mathcal{T}_{\q} (E)
\label{Fhh2}
\eeqa
where 
\beq  \mathcal{T}_{\q} (\hbar\omega)=2\frac{\im{g_{\rm w}^{\rm R}(\q,\omega)}^2e^{-2qd}}{|1-g_{\rm w}^{\rm R}(\q,\omega)^2e^{-2qd}|^2}
\label{T}
\eeq
is the dimensionless transmission coefficient. We thus recover on rigorous grounds the Landauer-like formula that was suggested in the main text.  

To obtain the scaling of the driving force with $d$, we make the same approximations as in the previous subsection. Namely, we neglect the denominator of Eq. \eqref{T}, the momentum dependence of the water surface response function and approximate 
\beq \im{g_{\rm w}^{\rm R}(\q,\omega)}\approx \alpha_{\rm w}\frac{\omega_0\omega}{\omega_0^2+\omega^2}\eeq
where $\omega_0\approx 0.3$ THz $\ll T/\hbar$ is the effective Debye frequency and $\alpha_{\rm w}$ an amplitude factor. 
Still to first order in velocities, we then obtain 
\beq
\frac{\tb{F}_{\rm hh}}{\A}\approx\frac{\alpha_{\rm w}T\Delta\v}{2\pi^2 }\int_0^\infty\dd q\, q^3e^{-2qd} \int_0^\infty\dd\omega\frac{\omega_0^2}{(\omega_0^2+ \omega^2)^2}=\frac{3}{64\pi^2}\frac{\alpha_{\rm w}T}{ \omega_0 d^4}\Delta\v \propto \frac{1}{d^4}
\label{d4}
\eeq 
where $\Delta\v=\v_{\rm A}-\v_{\rm B}$. We thus recover from rigorous theory the $1/d^4$ scaling obtained from the simple model in the main text. 

We may thus introduce the hydron-hydron friction coefficient as $\lambda_{\rm hh} = \F_{\rm hh} / (\mathcal{A} \Delta \v)$. $\F_{\rm hh}$ is the driving force for the flow tunneling effect: it induces the flow of fluid $B$ in response to the flow of fluid $A$. In the steady state, $\F_{\rm hh}$ is balanced by the classical (roughness-induced) friction $\F_{\rm cl} = - \lambda_{\rm cl} \mathcal{A} \v_{\rm B}$ exerted on fluid $B$ by the solid wall, so that
\beq 
v_{\rm B}=\frac{\lambda_{\rm hh}}{\lambda_{\rm hh}+\lambda_{\rm cl}}v_{\rm A}.
\eeq
This result is compared to MD simulations in Fig. \ref{SI_fig5}. When $\lambda_{\rm hh}$ is computed with Eq.~\eqref{Fhh2}, we obtain quantitative agreement with the simulations; Eq.~\eqref{d4} matches the long-distance scaling.

\subsection{Flow tunneling through an active solid}

\begin{figure*}
 \centering \includegraphics[width=0.95\textwidth]{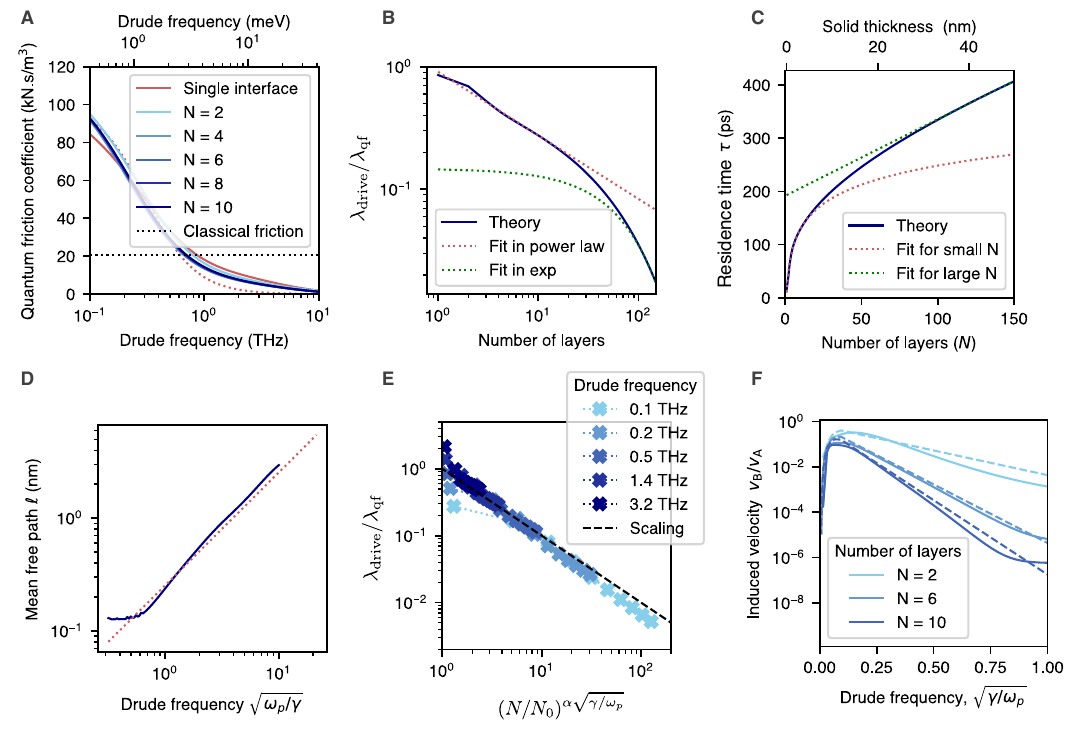} 
 \caption{\textbf{Scaling regimes.}
 \tb{(A)} Quantum friction coefficient of a water slab on a layer Drude oscillators (in red) and on $N$ layers of Drude oscillators with another slab of water behind. The red dashed line corresponds to the simple estimate of quantum friction Eq. \eqref{QF_simple_estimate}. The black dashed line corresponds to the classical friction. Here, the quantum friction is plotted as a function of the Drude frequency $\omega_p$ with a relaxation rate of $\gamma=10^{-2}$ THz. 
  \tb{(B)} Ratio $\lambda_{\rm drive}/\lambda_{\rm qf}$ as a function of the number of layers with a Drude frequency $\omega_p$=1.4 THz and a relaxation rate of $\gamma=10^{-2}$ THz. The dashed red curve is an algebraic fit and the dashed green curve is an exponential fit. 
\tb{(C)} Residence time $\tau$ as defined in Eq. \eqref{def_tau} as a function of the number of layers with a Drude frequency $\omega_p=1.4$ THz and a relaxation rate of $\gamma=10^{-2}$ THz. The dashed red curve is a fit in $\log{N}$ valid at small distances (algebraic regime) while the dashed green curve is a linear fit valid at large distances (exponential regime). 
  \tb{(D)} Hydron mean free path $\ell$, as defined in Eq.~\eqref{tau}, versus rescaled Drude frequency $\sqrt{\omega_p/\gamma}$ at $ \gamma = 10^{-1}~\rm THz$. The dashed line is $\ell = \ell_0 \sqrt{\omega_p/\gamma}$.
  \tb{(E)} Ratio $\lambda_{\rm drive}/\lambda_{\rm qf}$ versus rescaled number of layers. At not too large $N$, the all values of $\omega_p$ collapse onto the scaling expression of Eq.~\eqref{algebraic}. We have fixed $\gamma = 10^{-2}~\rm THz$. 
  \tb{(F)} Tunneling efficiency $v_B/v_A$ versus rescaled Drude frequency $\sqrt{\gamma/\omega_p}$. The dashed lines correspond to Eq.~\eqref{exp_scaling}. We have fixed $\gamma = 10^{-2}~\rm THz$. 
 } 
 \label{SI_fig6}
\end{figure*}

Let us now study the flow tunneling in presence of the solid's excitation modes. For this, we compute the out-of-equilibrium friction force between the solid and the induced flow.
To first order in velocities, this friction force is of the form 
\beq \frac{\tb{F}_{N\rightarrow B}}{\A}=\lambda_{\rm drive}\v_{\rm A}- \lambda_{\rm qf}\v_{\rm B} \label{drive_qf}\eeq
where the first term corresponds to the driving force due to the imposed flow and the second term corresponds to dissipation due to the quantum friction experienced by the induced fluid flowing at velocity $\v_{\rm B}$. Let us stress, however, that these two terms split only to first order in velocities, and they both originate in the same fluctuating Coulomb interactions between the liquid and the solid. 

In practice $\tb{F}_{N \rightarrow B}$ is computed by numerically carrying out the integrals in Eq.~\eqref{generic_friction} (with surface response functions renormalized through the iterative procedure described in Sec. III D), for imposed values of $\v_{\rm A}$ and $\v_{\rm B}$ in the linear response regime. By evaluating $\tb{F}_{N \rightarrow B}$ for two pairs of velocities $(\v_{\rm A}, \v_{\rm B}^1)$ and $(\v_{\rm A}, \v_{\rm B}^2)$, we obtain the friction coefficients $\lambda_{\rm drive}$ and $\lambda_{\rm qf}$. 

In addition to the flow tunneling driving force $\tb{F}_{N \rightarrow B}$, fluid $B$ is subject to classical friction on the $N^{\rm th}$ solid layer ($\tb{F}_{\rm cl} = - \lambda_{\rm cl} \A \v_{\rm B}$), and to the direct friction force with fluid $A$ if the solid is very thin ($ \tb{F}_{\rm hh}=\lambda_{\rm hh} \A(\v_{\rm A}-\v_{\rm B})$). Finally, imposing momentum balance in the steady state yields the magnitude of flow tunneling: 
\beq  \v_{\rm B}=\frac{\lambda_{\rm drive}+\lambda_{\rm hh}}{\lambda_{\rm qf}+\lambda_{\rm cl}+\lambda_{\rm hh}}\v_{\rm A}. \eeq
For $N > 1$, we may neglect $\lambda_{\rm hh}$, so that
\beq  \frac{v_{\rm B}}{v_{\rm A}}\approx\frac{\lambda_{\rm qf}}{\lambda_{\rm qf}+\lambda_{\rm cl}}\frac{\lambda_{\rm drive}}{\lambda_{\rm qf}}. \eeq

\subsection{Scaling regimes of the tunneling efficiency}

The dependence of the tunneling efficiency on the number of solid layers $N$ is contained in the dimensionless ratio $\lambda_{\rm drive}/ \lambda_{\rm qf}$. This ratio is plotted as a function of $N$ (for representative values of $\omega_p$ and $\gamma$) in Fig.~\ref{SI_fig6}B. It displays a crossover from a power law scaling at small $N$ to an exponential scaling at large $N$. As in the main text, we may define a "hydron residence time"
\beq \tau=-\frac{\log(\lambda_{\rm drive}/\lambda_{\rm qf})}{\gamma}.
\label{def_tau}\eeq
The exponential scaling for $\lambda_{\rm drive}/\lambda_{\rm qf}$ translates into a linear scaling for $\tau$ (Fig.~\ref{SI_fig6}C), anticipated on qualitative grounds in the main text: 
\beq
\tau \approx \tau_{\rm sl} + \frac{N d_0}{\ell} = \tau_{\rm sl} + \frac{d}{\ell}, 
\label{tau}
\eeq
which defines the hydron mean free path $\ell(\omega_p, \gamma)$. We find numerically that the mean free path scales as $\ell\approx\ell_0\sqrt{\frac{\omega_p}{\gamma}}$ with $\ell_0\approx 0.26$ nm (Fig. \ref{SI_fig6}D). Altogether, the tunneling efficiency in this regime is given by 
\beq
\frac{v_{\rm B}}{v_{\rm A}} \approx \frac{\lambda_{\rm qf}}{\lambda_{\rm qf} + \lambda_{\rm cl}} e^{- \gamma \tau_{\rm sl} + (d/\ell_0) \sqrt{\gamma/\omega_p}}. 
\label{exp_scaling}
\eeq
The power law regime is described by 
\beq
 \frac{\lambda_{\rm drive}}{\lambda_{\rm qf}}\approx\left(\frac{N_0}{N}\right)^{\alpha\sqrt{\gamma/\omega_p}}
 \label{algebraic}
 \eeq
with $N_0\approx 0.84$ and $\alpha\approx 6.2$. Fig. \ref{SI_fig6}E shows the collapse of $\lambda_{\rm drive}/ \lambda_{\rm qf}$ on this scaling form for a range of values of $\omega_{\rm p}$, at small enough $N$. 
Nevertheless, we find that, even at not too large $N$, Eq.~\eqref{exp_scaling} still provides a reasonable description of the tunneling efficiency $v_{\rm B}/v_{\rm A}$, as shown in Fig. \ref{SI_fig6}F. Eq.~\eqref{exp_scaling} may therefore be used to qualitatively discuss the influence of the parameters $N$, $\gamma$ and $\omega_p$ on the tunneling efficiency.

\bibliography{main_bibliography.bib}